\documentclass[final]{IEEEtran}

\usepackage[usenames,dvipsnames]{xcolor}
\usepackage{amsfonts}
\usepackage{cite} 
\usepackage{amsmath} 
\usepackage{amssymb}  
\usepackage{graphicx}
\usepackage{bm}
\usepackage{colortbl}
\usepackage{epsfig}
\usepackage{epstopdf}
\usepackage[normal]{subfigure}
\usepackage{mathtools}
\usepackage{etoolbox}
\usepackage{lipsum}
\usepackage{caption}
\usepackage{soul}
\usepackage[colorlinks=true]{hyperref}
\usepackage{verbatim}

\usepackage{comment}
\usepackage{algorithm,algorithmic}
\newenvironment{varalgorithm}[1]
  {\algorithm\renewcommand{\thealgorithm}{#1}}
  {\endalgorithm}

\def\IID{\mathop{\mathrm{i.i.d.}}}  
\def\diag{\mathop{\mathrm{diag}}}   
\def\trace{\mathop{\mathrm{trace}}} 
\def\rank{\mathop{\mathrm{rank}}}    
\def\rdf{\mathop{\mathrm{RDF}}}     
\def\mse{\mathop{\mathrm{MSE}}}     
\def\rvs{\mathop{\mathrm{RVs}}}     
\def\rv{\mathop{\mathrm{RV}}}       
\def\abs{\mathop{\mathrm{abs}}}     
\def\ind{\mathop{\mathrm{in}}}     

\newtheorem{theorem}{\bf Theorem}
\newtheorem{lemma}{\bf Lemma}
\newtheorem{definition}{\bf Definition}
\newtheorem{corollary}{\bf Corollary}

\newtheorem{remark}{\bf Remark}

\newtheorem{proposition}{\bf Proposition}
\newtheorem{example}{\bf Example}

\newcommand{\T}{^{\mbox{\tiny T}}}

\begin{document}

\title{{Indirect NRDF for Partially Observable Gauss-Markov Processes with MSE Distortion: Complete Characterizations and Optimal Solutions}}

\author{Photios A. Stavrou and Mikael Skoglund
\thanks{The authors have received funding by the KAW Foundation and the Swedish Foundation for Strategic Research.}
\thanks{The authors are with the Division of Information Science and Engineering, KTH Royal Institute of Technology, Sweden {\it email: \tt \{fstavrou,skoglund\}@kth.se}.}
}

\maketitle

\begin{abstract}
{In this paper we study the problem of characterizing and computing the nonanticipative rate distortion function (NRDF) for partially observable multivariate Gauss-Markov processes with hard mean squared error (MSE) distortion constraints. For the finite time horizon case, we first derive the complete characterization of this problem and its corresponding optimal realization which is shown to be a linear functional of the current time sufficient statistic of the past and current observations signals. We show that when the problem is strictly feasible, it can be computed via semidefinite programming (SDP) algorithm. For time-varying scalar processes with average total MSE distortion we derive an optimal closed form expression by means of a dynamic reverse-waterfilling solution that we also implement via an iterative scheme that convergences linearly in finite time, and a closed-form solution under pointwise MSE distortion constraint. For the infinite time horizon, we give necessary and sufficient conditions to ensure that asymptotically the sufficient statistic process of the observation signals achieves a steady-state solution for the corresponding covariance matrices and impose conditions that allow existence of a time-invariant solution. Then, we show that when a finite solution exists in the asymptotic limit, it can be computed via SDP algorithm. We also give strong structural properties on the characterization of the problem in the asymptotic limit that allow for an optimal solution via a reverse-waterfilling algorithm that we implement via an iterative scheme that converges linearly under a finite number of spatial components. Subsequently, we compare the computational time needed to execute for both SDP and reverse-waterfilling algorithms when these solve the same problem to show that the latter is a scalable optimization technique. Our results are corroborated with various simulation studies and are also compared with existing results in the literature.} 
\end{abstract}

{
}
\section{Introduction}

\par Nonanticipatory $\epsilon-$entropy was introduced in
\cite{gorbunov:1972,gorbunov:1972b} motived by real-time communication with minimal encoding and decoding delays. 
This entity is shown to be a tight lower bound on causal codes for scalar processes \cite{derpich:2012} whereas for vector processes it provides a tight lower bound at high rates on causal codes and on the average length of all causal prefix free codes \cite{stavrou:2018} (also termed zero-delay coding). 
\par Inspired by the usefulness of nonanticipatory-$\epsilon$ entropy in real-time communication, Tatikonda {\it et al.} in \cite{tatikonda:2004} reinvented the same measure under the name sequential rate distortion function ($\rdf$)\footnote{In the literature this information measure can also be found under the name nonanticipative $\rdf$ (NRDF) \cite{charalambous:2014}.} to study a linear fully observable Gaussian closed-loop control system over a memoryless communication channel subject to rate constraints. In particular, the authors of \cite{tatikonda:2004} used the sequential $\rdf$ subject to a pointwise $\mse$ distortion constraint to describe a lower bound on the minimum cost of control for scalar-valued Gaussian processes and a suboptimal lower bound for the multivariate case obtained by means of a reverse-waterfilling algorithm \cite[10.3.3]{cover-thomas:2006}.\footnote{The suboptimality of the lower bound obtained in \cite{tatikonda:2004} for multivariate Gaussian processes was recently identified in  \cite{kostina:2019a,stavrou:2020tac}.}
\par Tanaka {\it et al.} in \cite{tanaka:2017} revisited the
estimation/communication part of the problem introduced by Tatikonda {\it et al.} and showed that the specific description of the sequential $\rdf$ is semidefinite representable. Around the same time, Stavrou {\it et al.} in \cite{stavrou:2018siam} solved the general KKT conditions that correspond to the rate distortion characterization of the optimal estimation problem in \cite{tatikonda:2004} and proposed a dynamic reverse-waterfilling characterization (for both pointwise and total $\mse$ distortions) that computes optimally the KKT conditions as long as all dimensions of the multidimensional setup are active, which is the case at high rates regime. In addition, in \cite{stavrou:2018siam} they found the optimal linear coding policies (by means of a linear forward test-channel realization) that achieve the specific rate distortion characterization thus filling a
gap created in \cite[Theorem 5]{gorbunov:1972}. Recently, the optimal realization therein was used as a benchmark in \cite{fuglsig:2019} to derive bounds on a zero delay multiple description source coding problem with feedback.
\par Kostina and Hassibi in \cite{kostina:2019a} revisited the framework of \cite{tatikonda:2004} {and derived bounds on the optimal rate-cost tradeoffs in control} for time-invariant fully observable multivariate Markov processes {under the assumption of uniform cost (or distortion) allocation}. Recently, Charalambous {\it et al.} in \cite{charalambous:2019cdc} used a state augmentation technique to extend the characterization of the Gaussian nonanticipatory $\epsilon-$entropy derived in \cite{gorbunov:1972b} to nonstationary multivariate Gaussian autoregressive models of any finite order.

\par The extension of the framework of \cite{tatikonda:2004} to stochastic linear partially observable Gaussian control systems under noisy or noiseless communication channels was initially studied in \cite{tanaka:2018} {whereas a variation of the uncontrolled problem is studied in \cite{tanaka:2015}}. {Particularly, Tanaka in \cite{tanaka:2015} considered the estimation/communication part of the problem and derived performance limitations by minimizing a sequential $\rdf$ with {\it soft weighted pointwise MSE distortion constraints}. To deal with this problem, he first reduced the time-varying partially observable Gaussian system into a fully observable one by employing a pre-Kalman filtering (pre-KF) algorithm. Then, he assumed {\it \'a priori} a structural result on its observations process to ensure the invertibility of the pre-KF algorithm and hence to guarantee that the {\'a posteriori} state estimate between the state process and the observations process computed by the pre-KF is an information lossless operation of the true observations process at each instant of time. Armed with this result and a modified MSE distortion constraint he then showed that the resulting problem can be equivalently reformulated as {fully observable multi-letter optimization for which a cascade realization} was proposed via the connection of a pre-KF, a covariance scheduling semidefinite programming (SDP) algorithm, an additive white Gaussian noise (AWGN) channel and a post-KF algorithm. The stationary case of the specific optimization problem is also briefly discussed. Tanaka {\it et al.} in \cite[Section VII, Eq. (38)]{tanaka:2018} considered a multi-letter optimization problem via directed information \cite{massey:1990} between the observations process and the controlled process and an average total hard constraint obtained via the classical LQ cost. Again the major result therein is a modified fully observable multi-letter optimization problem for controlled processes achieved by a cascade realization in the spirit of \cite{tanaka:2015} only for the finite time horizon problem.} 
\par {Despite the interesting analysis of \cite{tanaka:2015,tanaka:2018}, there are {\it several important open questions} still unanswered even for the estimation/communication problem. For instance, in both \cite{tanaka:2015,tanaka:2018} it is not clear what is the complete characterization that needs to be solved similar to what is already known for example when the input data are modeled via a linear fully-observable multidimensional system driven by additive white Gaussian noise (see, e.g., \cite[Eq. (5.22)]{stavrou:2018siam}). Moreover, the complete (minimum) realization of the optimal test-channel distribution including the identification of the reverse-waterfilling parameters that achieve the specific characterization is also missing. Another important question has to do with the conditions that are needed to ensure (strict) feasibility of the optimization problem in both finite and infinite time horizon. Equally important questions include the derivation of optimal or suboptimal (numerical or analytical) solutions for this problem for both scalar or beyond scalar processes as well as the analysis of the problem for high dimensional systems that necessitates scalable optimization algorithms (an issue already known from the analysis of \cite{stavrou:2021tac}).}
\par {Kostina and Hassibi in \cite{kostina:2019a} understood some of the previous questions and derived analytical bounds on the exact solutions of the estimation and control problems for  time-invariant multivariate jointly Gaussian processes {\it again under the assumption of uniform distortion allocation}. Hence, a natural open question related to the bounds in \cite{kostina:2019a} is their tightness for multidimensional systems. This question is also related to the fact that {\it no insightful examples appeared in the literature so far to compute optimally partially observable multivariate Gauss-Markov processes} and compare with the closed form bounds obtained in \cite{kostina:2019a}.}

\subsection{Contributions}\label{subsec:contributions}

\par {In this work we study the problem of characterizing and computing the NRDF (hereinafter termed indirect NRDF) for partially observable multivariate Gauss-Markov processes under hard MSE distortion
constraints in both finite and infinite time horizon.  We obtain the following major results.\\
{\bf (R1)} We derive the complete characterization of the  indirect NRDF for a partially observable time-varying Gauss-Markov process with an average total or pointwise MSE distortion constraint and we completely specify the corresponding optimal test-channel realization which is a linear functional of the current sufficient statistic of the past and present observation signals (see Theorem \ref{theorem:complete_characterization_suff_stat});\\
{\bf (R2)} We give sufficient conditions to ensure existence of a finite solution for any fixed finite time horizon (see Remark \ref{remark:suff_cond_characterization}) and show that the problem for time-varying multivariate Gaussian processes is semidefinite representable (see Theorem \ref{theorem:numer_col_finite_time});\\
{\bf (R3)} For time-varying scalar processes under average total MSE distortion constraints we derive the optimal closed form solution via a dynamic reverse-waterfilling algorithm (see Theorem \ref{theorem:opti_num_sol_scalar_tv}) that we implement in Algorithm \ref{algo1} whereas for pointwise MSE distortion constraints we derive the optimal closed form solution (see Corollary \ref{corollary:closed_form_scalar_tv});\\
{\bf (R4)} For the infinite time horizon, we restrict our problem to time-invariant processes and identify necessary and sufficient conditions (i.e., detectability and stabilizability of appropriate pair of matrices) to ensure a steady state solution of the error covariance matrices of the sufficient statistic process (see Lemma \ref{lemma:necessary_suff}) and then we give conditions that allow for a time-invariant characterization in the asymptotic limit (see Theorem \ref{theorem:existence_asymptotic_limit});\\
{\bf (R5)} For the infinite time horizon, we show that when a  finite solution exists the problem is semidefinite representable (see Corollary \ref{corollary:complete_optim_numer_infinite_horizon}) and under certain strong structural properties on the asymptotic characterization of the problem (see Proposition \ref{proposition:strong_structural_properties}) we derive an optimal scalable reverse-waterfilling solution (see Theorem \ref{theorem:rev_water}) with its algorithmic embodiment (see Algorithm \ref{algo2});\\
{\bf (R6)} We supplement our major results with numerical validations including connections with \cite{kostina:2019a} (see Section \ref{sec:num_sim}).
}
{\noindent{\it Additional results and comparison to prior art.} To be able to prove the major results {\bf (R1)}-{\bf (R5)}, we first prove the exact expression of the lower bound that needs to be studied when the low-delay source coding system is modeled by partially observable Gauss-Markov process (see Definition \ref{def:lower_bound_exact}). Then, we show via a modification of the distortion constraint that the specific information measure can be reduced to the classical NRDF \cite{gorbunov:1972,stavrou:2018siam} hence it has similar functional and topological properties, i.e., convexity, lower-continuity etc. For jointly Gaussian processes we use a pre-KF algorithm (similar to \cite{tanaka:2015}) to prove structural properties via a sufficient statistic approach together with a data processing inequality (see Lemma \ref{lemma:suff_stat_gaussian}) that result into the same expression of the multi-letter optimization first appeared in \cite[Eq. (17)]{tanaka:2015} with {\it hard average total} MSE distortion constraints instead of soft pointwise MSE distortion constraints that were assumed in \cite{tanaka:2015} for both finite and infinite time horizon (see Definition \ref{def:remote_nrdf_gaussian}). Then, we apply \cite[Theorem 4.1]{stavrou:2018siam} and prove that the multi-letter optimization problem of Definition \ref{def:remote_nrdf_gaussian} can be simplified to a single-letter sequential optimization problem in which we only need the current sufficient statistic of the past and present observations symbols (see Proposition \ref{proposition:structural_result}). The computational complexity of the SDP algorithm in finite time horizon is discussed in Remark \ref{remark:ontheorem_numer_sol_finite_time} and this analysis also includes the single stage case. The computational time complexity and convergence of Algorithm \ref{algo1} is analyzed in Remark \ref{remark:complexity:algo1} and that of Algorithm \ref{algo2} in Remark \ref{remark:compl_algo_2}. Note that Theorem \ref{theorem:rev_water} and its algorithmic embodiment Algorithm \ref{algo2} are extremely important for two reasons; first we can gain better insights of the problem in the infinite time horizon (for instance it paves the way for one to derive optimal closed form solutions beyond scalar processes thus generalizing similar results obtained for the special case of fully observable time-invariant multivariate Gauss-Markov processes studied recently in \cite[Section IV]{stavrou:2021tac}) and, second, Algorithm \ref{algo2} as Table \ref{table:comparison} suggests can operate much faster than the SDP algorithm in high dimensional systems (it is scalable). Our numerical simulation in Example \ref{example:2} apart from verifying numerically that both Corollary \ref{corollary:complete_optim_numer_infinite_horizon} and Theorem \ref{theorem:rev_water} coincide under certain structural properties, it also shows that the corresponding analytical lower bound obtained for partially observable time-invariant multidimensional Gauss-Markov processes via \cite[Corollary 1, Theorem 9]{kostina:2019a} is not tight in general but a fairly tight performance (not exact) can be observed at very low distortion. Consequently, its utility to controlled processes in \cite[Theorem 5]{kostina:2019a} should be seen under this consideration. Example \ref{example:3}, shows the utility of Algorithm \ref{algo1} when we restrict our system to time-invariant scalar processes, namely, for certain necessary and sufficient conditions on the pre-KF algorithm it recovers the steady-state solution of Corollary \ref{corollary:scalar_time_inv}. Finally, for every result in this paper we recover or explain how to recover as a special case the corresponding results obtained for fully-observable Gauss-Markov processes. 
}

{\bf Notation.} We let $\mathbb{R}=(-\infty,\infty)$, $\mathbb{Z}$=$\{\ldots,-1,0,1,\ldots\}$, $\mathbb{N}_0=\{0,1,\ldots\}$, $\mathbb{N}_0^n=\{0,1,\ldots,n\}$,~$n\in\mathbb{N}_0$. Let  ${\cal X}$ be a finite dimensional Euclidean space and ${\cal B}({\cal X})$ the Borel $\sigma$-field of ${\cal X}$. A random variable ($\rv$) defined on some probability space $(\Omega, \mathcal{F}, {\mathbb P})$ is a map ${\bf x}: \Omega \longmapsto \mathcal{X}$, where $(\mathcal{X}, \mathcal{B}(\mathcal{X}))$ is a measurable space. We denote a sequence of $\rvs$ by ${\bf x}_r^t \triangleq ({\bf x}_r, {\bf x}_{r+1}, \ldots,{\bf x}_t), (r, t) \in {\mathbb Z}\times {\mathbb Z}, t\geq r$, and their realizations by ${x}_r^t \in  {\mathcal{X}}_r^t \triangleq \times_{k=r}^t {\mathcal{X}}_k$, for simplicity. If $r=-\infty$ and $t=-1$, we use the notation ${\bf x}_{-\infty}^{-1}={\bf x}^{-1}$, and if $r=0$,  we use the notation ${\bf x}_0^t = {\bf x}^t$. The distribution of the $\rv$ ${\bf x}$ on $\mathcal{X}$ is denoted by ${\bf P}(dx)$. The conditional distribution of a ${\rv}$ ${\bf y}$ given ${\bf x}=x$ is denoted by ${\bf P}(dy|x)$. {The transpose and covariance of a random vector ${\bf x}$ are denoted by ${\bf x}\T$ and $\Sigma_{\bf x}$. {We denote the determinant, trace, rank, diagonal, diagonal elements, and eigenvalues of a square matrix $S\in\mathbb{R}^{p\times{p}}$ by $|S|$, $\trace(S)$, $\rank(S)$, $\diag(S)$, $[\cdot]_{ii}$ and $\{\mu_{S,i}\}_{i=1}^p$ and $S^\dagger$. We denote the transpose and the pseudo-inverse of a real (rectangular) matrix $F\in\mathbb{R}^{p\times{m}}$ by $F\T$ and $F^\dagger$.} The notation $\Sigma\succ{0}$ (resp. $\Sigma\succeq{0}$) denotes a positive definite (resp. positive semi-definite) matrix. The notation $A\succ{B}$ (resp. $A\succeq{B}$) means $A-B\succ{0}$ (resp. $A-{B}\succeq{0}$).  We denote a $p\times{p}$ identity matrix by $I_p$.}  
${R}^{G}(D)$ denotes the Gaussian version of the $\rdf$. The expectation operator is denoted by $\mathbb{E}\{\cdot\}$; $||\cdot||$ denotes Euclidean norm; $[\cdot]^{+}\triangleq\max\{0,\cdot\}$. {We denote by $\abs(|\cdot|)$ the absolute value of a determinant.}  

\section{Problem statement}\label{sec:problem_statement}

\begin{figure}[htp]
\centering
\includegraphics[width=\columnwidth]{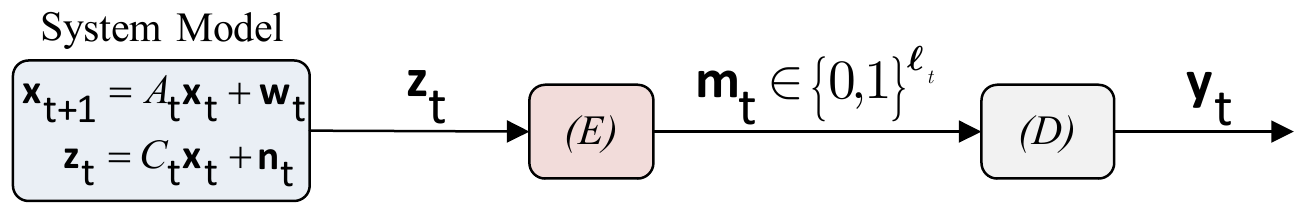}
\caption{Causal coding of partially observable Gauss-Markov process.}\label{fig:partially_observable}
\end{figure}

We consider the causal source coding setup of Fig.~\ref{fig:partially_observable}. In this setting, the ``hidden'' $\mathbb{R}^p$-valued source is modeled by a {discrete-time time-varying partially observable Gauss-Markov process as follows
\begin{align}
{\bf x}_{t+1}&=A_t{\bf x}_{t}+{\bf w}_t,~{\bf x}_0=\bar{x},\label{state_process_tv}\\
{\bf z}_{t}&=C_t{\bf x}_{t}+{\bf n}_{t},~t\in\mathbb{N}_0,\label{observation_process_tv}
\end{align}
where $A_t\in\mathbb{R}^{p\times{p}}$ is a square non-random matrix, $C_t\in\mathbb{R}^{m\times{p}}$ is a rectangular non-random matrix with $m\leq{p}$, ${\bf x}_0\in\mathbb{R}^p\sim(0;\Sigma_{{\bf x}_0})$, $\Sigma_{{\bf x}_0}\succ{0}$ is the initial state, ${\bf w}_t\in\mathbb{R}^p\sim{\cal N}(0;\Sigma_{{\bf w}_t})$, $\Sigma_{{\bf w}_t}\succ{0}$ is an independent sequence,  ${\bf n}_t\in\mathbb{R}^m\sim{\cal N}(0;\Sigma_{{\bf n}_t})$, $\Sigma_{{\bf n}_t}\succeq{0}$, is an independent sequence, independent of $\{{\bf w}_t:~t\in\mathbb{N}_0\}$, whereas ${\bf x}_0$ is independent of $\{({\bf w}_t,{\bf n}_t):~t\in\mathbb{N}_0\}$.} 
\paragraph*{System's operation} {At every time instant, the {\it encoder} observes the impair measurement ${\bf z}_t$ (provided ${\bf z}^{t-1}$ are already observed) and generates the data packet ${\bf m}_t\in{\cal M}_t\subset\{0,1\}^{\ell_t}$ of instantaneous expected rate $R_t=\mathbb{E}|{\bm \ell}_t|$, where $|{\bm \ell}_t|$ denotes the binary sequence of ${\bm \ell}_t$. At time $t$, ${\bf m}_t$ is transmitted across a noiseless channels with rate $R_t$. Upon receiving ${\bf m}^t$, a {\it minimum MSE (MMSE) decoder} immediately produces an estimate ${\bf y}_t$ of the source sample {${\bf x}_t$, under the assumption that ${\bf y}^{t-1}$ are already reproduced}. We assume that at time $t=0$ there is no prior information whereas the clocks of the encoder and the decoder are synchronized. 
\begin{align}
\begin{split}
{\cal (E)}:~m_t&=f_t(m^{t-1},{z}^t),~m^{-1}=\emptyset, z^{-1}=\emptyset,\\
{\cal (D)}:~y_t&=g_t(m^t).
\end{split}
\end{align}\label{coding_functions} 
}
{
\noindent{\it Distortion Constraint}. The distortion constraint is described by the {\it average total $\mse$ distortion} given by:
\begin{align}
\frac{1}{n+1}\sum_{t=0}^n\mathbb{E}\left\{||{\bf x}_t-{\bf y}_t||^2\right\}\leq{D},\label{distortion_constraint}
\end{align}
and its asymptotic limit by
\begin{align}
\limsup_{n\longrightarrow\infty}\frac{1}{n+1}\sum_{t=0}^n\mathbb{E}\left\{||{\bf x}_t-{\bf y}_t||^2\right\}\leq{D}.\label{distortion_constraint_infinite_time}
\end{align}
{\it Performance.} The performance of the multi-input multi-output (MIMO) system in Fig. \ref{fig:partially_observable} after some finite  $n$ can be cast by the following optimization problem:
\begin{align}
R^{c}_{[0,n],\ind}(D)\triangleq\inf_{\substack{m_t=f_t(m^{t-1},z^t),~t\in\mathbb{N}_0^n\\~y_t=g_t(m^t)\\
\mbox{Eq. ~\eqref{distortion_constraint}}}}\frac{1}{n+1}\sum_{t=0}^nR_t.
\label{def:operational_causal_finite_time}
\end{align}
The asymptotic limit of \eqref{def:operational_causal_finite_time} is given as follows:
\begin{align}
R^{c}_{\ind}(D)\triangleq\inf_{\substack{m_t=f_t(m^{t-1},z^t),~t\in\mathbb{N}_0\\~y_t=g_t(m^t)\\
\mbox{Eq. ~\eqref{distortion_constraint_infinite_time}}}}\limsup_{n\longrightarrow\infty}\frac{1}{n+1}\sum_{t=0}^nR_t.
\label{def:operational_causal_infinite_time}
\end{align}
}

%
%
%
{
\section{The exact lower bound on \eqref{def:operational_causal_finite_time}}
\label{sec:lower_bound_exact}
}
\par {In this section, we first prove the exact lower bound that corresponds to the operational rates given in \eqref{def:operational_causal_finite_time}, \eqref{def:operational_causal_infinite_time} because its proof or construction analysis is to the best of the authors' knowledge not included in \cite{tanaka:2015,tanaka:2018,kostina:2019a} or elsewhere.}

{We start by writing the data processing of information for the MIMO system of Fig. \ref{fig:partially_observable} in terms of its joint distribution.  In particular, the joint distribution induced by the joint process  $\{({\bf z}_t, {\bf m}_t, {\bf y}_t):~t\in\mathbb{N}_0^n\}$ admits the following decomposition:
\begin{align}
&{\bf P}(dy^n,dm^n,dz^n)=\otimes_{t=0}^n{\bf P}(dy_t,dm_t,dz_t|y^{t-1},m^{t-1},z^{t-1})\nonumber\\
&=\otimes_{t=0}^n{\bf P}(dy_t|y^{t-1},z^t,m^t)\otimes{\bf P}(dm_t|m^{t-1},z^t,y^{t-1})\nonumber\\
&\qquad\otimes{\bf P}(dz_t|z^{t-1},y^{t-1},m^{t-1})\nonumber\\
&\stackrel{(a)}=\otimes_{t=0}^n{\bf P}(dy_t|y^{t-1},z^{t},m^t)\otimes{\bf P}(dm_t|m^{t-1},z^t,{y}^{t-1})\nonumber\\
&\qquad\otimes{\bf P}(dz_t|z^{t-1}),\label{joint_data processing}
\end{align}
where $(a)$ stems from the fact that we assume in our system the following natural conditional independence constraints
\begin{align}
{\bf P}(dz_t|z^{t-1},y^{t-1},m^{t-1})&={\bf P}(dz_t|z^{t-1}),\label{cond_independence_constr_1}\\
{\bf P}(dy_t|y^{t-1},z^t,m^t)&={\bf P}(dy_t|y^{t-1},m^t).\label{cond_independence_constr_2}
\end{align}
\begin{remark}(Trivial initial information)
To be consistent to the setup of Fig. \ref{fig:partially_observable}, in \eqref{joint_data processing} we assume that the joint distribution ${\bf P}(dz^{-1},dm^{-1},d{y}^{-1})$ generates trivial information. 
\end{remark}
The following data processing result, provides the appropriate information measure that can be used to compute a lower bound on \eqref{def:operational_causal_finite_time}. 
\begin{lemma}(Data processing inequalities)\label{lemma:data_processing}
Under the decomposition of the joint distribution in \eqref{joint_data processing}, the communication system in Fig. \ref{fig:partially_observable} admits the following data processing inequalities:
\begin{align}
\begin{split}
I({\bf z}^n;{\bf y}^n)\stackrel{\bf (ii)}\leq{I}({\bf z}^n;{\bf m}^n||{\bf y}^{n-1})
\stackrel{\bf (i)}\leq\sum_{t=0}^nR_t,
\end{split}\label{dpi}
\end{align}
where 
\begin{align}
{I}({\bf z}^n;{\bf y}^n)&=\sum_{t=0}^nI({\bf z}^t;{\bf y}_t|{\bf y}^{t-1}),\label{causally_cond_MI}\\
{I}({\bf z}^n;{\bf m}^n||{\bf y}^{n-1})&=\sum_{t=0}^n{I}({\bf z}^t;{\bf m}_t|{\bf m}^{t-1},{\bf y}^{t-1}), \nonumber
\end{align}
and $I({\bf z}^t;{\bf y}_t|{\bf y}^{t-1})<\infty$, ${I}({\bf z}^t;{\bf m}_t|{\bf m}^{t-1},{\bf y}^{t-1})<\infty$,~$\forall{t}$.
\end{lemma}
\begin{IEEEproof}
The proof follows precisely similar steps to the proof of \cite[Theorem 1]{stavrou:2020entropy} thus we omit it.
\end{IEEEproof}
}

{Next, we show how to formally construct the information measure \eqref{causally_cond_MI}.}\\
{{\bf Observations Process.} The observations process $\{{\bf z}_t:~t\in\mathbb{N}_0^n\}$ induces the sequence of conditional distributions ${\bf P}(dz_t|z^{t-1}),~t\in\mathbb{N}_0$. At $t=0$ we assume that ${\bf P}(dz_0|z^{-1})={\bf P}(dz_0)$ and by Bayes' rule we obtain 
\begin{align}
{\bf P}(dz^n)=\otimes_{t=0}^n{\bf P}(dz_t|z^{t-1}).\label{observations_distribution} 
\end{align}
It should be noted  that for the system model \eqref{state_process_tv}, \eqref{observation_process_tv}, at each instant of time, the conditional distribution of ${\bf P}(dz_t|z^{t-1})$ depends on the {\it posterior distribution} of the hidden data ${\bf x}_t$ given all the past observation symbols ${\bf z}^{t-1}$ via
\begin{align}
{\bf P}(dz_t|z^{t-1})=\int_{{\cal X}_t}{\bf P}(dz_t|x_t){\bf P}(dx_t|z^{t-1}).\label{observ_prior}
\end{align}  
{\bf Reproduction or ``test-channel''.} The reproduction process ${\bf y}_t$ parametrized by ${\cal Y}^{t-1}\times{\cal Z}^t$ induces the sequence of conditional distributions known as test-channels as follows ${\bf P}(dy_t|y^{t-1},z^t),~t\in\mathbb{N}_0^n$. {At $t=0$, no initial state information is assumed, hence  ${\bf P}(dy_0|y^{-1},z^0)={\bf P}(dy_0|z_0)$.} The sequence of conditional distributions $\{{\bf P}(dy_t|y^{t-1},z^t):~t\in\mathbb{N}_0\}$ uniquely defines the family of conditional distributions on ${\cal Y}^{n}$ parametrized by $z^n \in {\cal Z}^{n}$, given by
\begin{align}
{\bf Q}(dy^n|z^n)\triangleq\otimes_{t=0}^n{\bf P}(dy_t|y^{t-1},z^t),\label{reproduction:distributions}
\end{align}
and vice-versa. From \eqref{observations_distribution} and \eqref{reproduction:distributions}, we can uniquely define the joint distribution of $\{({\bf z}_t,{\bf y}_t):~t\in\mathbb{N}_0^n\}$ by
\begin{align}
{\bf P}(dy^n,dz^n)={\bf P}(dz^n)\otimes{\bf Q}(dy^n|z^n).\label{joint_distribution}
\end{align}
In addition, from \eqref{joint_distribution}, we can define the ${\cal Y}^{n}-$marginal distribution ${\bf P}(dy^n)\triangleq\otimes_{t=0}^n{\bf P}(dy_t|y^{t-1})$, where 
\begin{align}
{\bf P}(dy_t|y^{t-1})=\int_{{\cal Z}^t}{\bf P}(dy_t|y^{t-1},z^t)\otimes{\bf P}(dz^t|y^{t-1}).\label{marginal_output}
\end{align}
}
\par {Given the above construction of distributions we obtain the following variant of directed information \cite{massey:1990}
\begin{align}
I({\bf z}^n;{\bf y}^n) &\stackrel{(a)}{=} \sum_{t=0}^n \mathbb{E}\left\{\log \left( \frac{d{\bf P}(\cdot|{\bf y}^{t-1},{\bf z}^t)}{d{\bf P}(\cdot|{\bf y}^{t-1})}({\bf y}_t)\right)\right\}\nonumber\\
&\stackrel{(b)}{=}\sum_{t=0}^n I({\bf z}^t; {\bf y}_t|{\bf y}^{t-1}),  \label{eq_s}
\end{align} 
where $(a)$ is due to chain rule of relative entropy using the Radon-Nykodym derivative \cite{dupuis-ellis:1997};  $(b)$ follows by definition.
\begin{definition}(Exact lower bounds on \eqref{def:operational_causal_finite_time}, \eqref{def:operational_causal_infinite_time})\label{def:lower_bound_exact} For a given observation processes $\{{\bf z}_t:~t\in\mathbb{N}_0^n\}$ that induces the conditional distribution \eqref{observ_prior}, the exact lower bound on \eqref{def:operational_causal_finite_time}, hereinafter called {\it remote} or {\it indirect NRDF}, subject to \eqref{distortion_constraint} is defined as follows
\begin{align}
&{R}_{[0,n],\ind}(D) \triangleq \inf_{\substack{{\bf P}(dy_t|y^{t-1},z^t):~t\in\mathbb{N}_0^n\\~\mbox{Eq. ~\eqref{distortion_constraint}}}}I({\bf z}^n;{\bf y}^n).\label{finite_time_remote_nrdf}
\end{align}
Moreover, its asymptotic expression that corresponds to a lower bound on \eqref{def:operational_causal_infinite_time} is given by
\begin{align}
&{R}_{\ind}(D)\triangleq\inf_{\substack{{\bf P}(dy_t|y^{t-1},z^t):~t\in\mathbb{N}_0\\ \mbox{Eq. ~\eqref{distortion_constraint_infinite_time}}}}\limsup_{n\longrightarrow\infty}\frac{1}{n+1}I({\bf z}^n;{\bf y}^n),\label{infinite_time_remote_nrdf}
\end{align}
provided the limit in \eqref{infinite_time_remote_nrdf} takes a finite value.
\end{definition}
}
\par {Next, we further analyze the information measure introduced in Definition \ref{def:lower_bound_exact} and discuss some of its most important properties.} 
\par {The name indirect or remote NRDF is adopted because the specific information measure can be seen as an extension to causal processes (with memory) of the remote or indirect RDF defined for $\IID$ memoryless processes  $\{({\bf x}_t, {\bf z}_t, {\bf y}_t):~t\in\mathbb{N}_0^n\}$ or random variables $({\bf x}, {\bf z}, {\bf y})$ in the context of non-causal coding see, e.g., \cite{dobrushin:1962,wolf-ziv:1970}, \cite[Chapters 3.5, 4.5]{berger:1971}, \cite{witsenhausen:1980}. Following for instance the approach in \cite{witsenhausen:1980}, one can transform the indirect NRDF of \eqref{finite_time_remote_nrdf} into a direct NRDF by creating a modified distortion constraint. For completeness, next we include such steps to obtain the amended distortion constraint.
\begin{align}
&\sum_{t=0}^n\mathbb{E}\{||{\bf x}_t-{\bf y}_t||\}=\sum_{t=0}^n\int_{{\cal X}^t\times{\cal Y}^t}||{x}_t-{y}_t||^2{\bf P}(dx^t,dy^t)\nonumber\\
&=\sum_{t=0}^n\int_{{\cal X}^t\times{\cal Z}^t\times{\cal Y}^t}||{x}_t-{y}_t||^2{\bf P}(dx^t,dz^t,dy^t)\nonumber\\
&\stackrel{(\star)}=\sum_{t=0}^n\int_{{\cal Z}^t\times{\cal Y}^t}{\bf P}(dz^t,dy^t)\int_{{\cal X}^t}||{x}_t-{y}_t||^2{\bf P}(dx^t|z^t)\nonumber\\
&\stackrel{(\star\star)}=\sum_{t=0}^n\int_{{\cal Z}^t\times{\cal Y}^t}{\bf P}(dz^t,dy^t)\int_{{\cal X}_t}||{x}_t-{y}_t||^2{\bf P}(dx_t|z^t)\nonumber\\
&\stackrel{(\star\star\star)}=\sum_{t=0}^n\mathbb{E}\{\hat{d}({\bf z}^t,{\bf y}_t)\},\label{amended_distortion}
\end{align}
where $(\star)$ follows due to the conditional independence constraint ${\bf P}(dy^t|z^t,x^t)={\bf P}(dy^t|z^t),~$for any $t=0,1,\ldots,n$; $(\star\star)$ follows by the system model \eqref{state_process_tv}, \eqref{observation_process_tv}, i.e., 
\begin{align}
{\bf P}(dx^t|z^t)=\frac{{\bf P}(dz_t|x_t){\bf P}(dx_t|z^{t-1})}{\int_{{\cal X}_t}{\bf P}(dz_t|x_t){\bf P}(dx_t|z^{t-1})}\equiv{\bf P}(dx_t|z^t),\label{structural_property}
\end{align} 
$(\star\star\star)$ follows if we define 
\begin{align}
\hat{d}({\bf z}^t,{\bf y}_t)\triangleq\int_{{\cal X}_t}||{x}_t-{y}_t||^2{\bf P}(dx_t|z^t).\nonumber
\end{align}
Hence, \eqref{finite_time_remote_nrdf} can be equivalently reformulated as follows
\begin{align}
&{R}_{[0,n],\ind}(D) \triangleq \inf_{\substack{{\bf P}(dy_t|y^{t-1},z^t):~t\in\mathbb{N}_0^n\\~\frac{1}{n+1}\sum_{t=0}^n\mathbb{E}\left\{\hat{d}({{\bf z}^t},{\bf y}_t)\right\}\leq{D}}}I({\bf z}^n;{\bf y}^n),\label{finite_time_remote_nrdf_modified}
\end{align}
which corresponds precisely to a direct NRDF. It is easy to show that \eqref{finite_time_remote_nrdf_modified} is convex with respect to the test channels $\{{\bf P}(dy_t|y^{t-1},z^t):~t\in\mathbb{N}_0^n\}$ following for instance \cite{charalambous:2016}. In addition, $R_{[0,n],\ind}(D)$ is monotonically non-increasing, convex with respect to $D$, continuous in $D\in(D^{\min},\infty)$ and if $R_{[0,n],\ind}(D^{\min})<\infty$, then it is continuous in $D\in[D^{\min},\infty)$. It is also well known that, $R_{[0,n],\ind}(D)$ achieves smaller rates if in addition to $\{({\bf x}_t, {\bf z}_t):~t\in\mathbb{N}_0^n\}$ being a jointly Gaussian process with the linear evolution of \eqref{state_process_tv}, \eqref{observation_process_tv}, the joint process $\{({\bf x}_t, {\bf z}_t, {\bf y}_t):~t\in\mathbb{N}_0^n\}$ is also Gaussian because then $I({\bf z}^n;{\bf y}^n)\geq{I}^G({\bf z}^n;{\bf y}^n)$ (that is, the Gaussian version of $I({\bf z}^n;{\bf y}^n)$) which in turn implies that $R_{[0,n],\ind}(D)\geq{R}^G_{[0,n],\ind}(D)$ (see, e.g., \cite[Theorem 1.8.6]{ihara1993}).}
{
\section{Complete characterization and optimal computational methods: finite time horizon}\label{sec:complete_char_compute_finite_time}
}
\par{In this section, we assume that the end-to-end system in Fig. \ref{fig:partially_observable} is jointly Gaussian and we completely characterize {\it for the first time} the exact lower bound \eqref{finite_time_remote_nrdf_modified} in finite time and provide conditions to ensure a strictly feasible solution for the problem for any finite $n$.} 
\par {To completely characterize the problem in finite time we use a two-step approach. As a first step, we employ a pre-KF to create the {\it MMSE estimator} of ${\bf x}_t$ given the past and current observation symbols ${\bf z}^t$ for any $t$. Then, using a sufficient statistic approach we show that the MMSE estimator of the pre-KF is in fact under certain conditions a {\it sufficient statistic at each time instant} of the process ${\bf y}_t$ parametrized by ${\cal Y}^{t-1}$ as it contains all the information about ${\bf z}^t$.
\begin{lemma}\label{lemma:classical_kf}(Classical KF) For the jointly Gaussian system model of \eqref{state_process_tv}, \eqref{observation_process_tv}, define the {\'a priori} and {\'a posteriori} state estimates as $\widehat{\bf x}_{t|t-1}\triangleq\mathbb{E}\{{\bf x}_t|{\bf z}^{t-1}\}$ and $\widehat{\bf x}_{t|t}\triangleq\mathbb{E}\{{\bf x}_t|{\bf z}^{t}\}$, respectively, and their corresponding error covariance matrices by\footnote{{For jointly Gaussian systems, the conditional covariance is equal to its unconditional version \cite{anderson:1979}.}}
\begin{align}
\Sigma_{t|t-1}^{\bf x}&\triangleq\mathbb{E}\{({\bf x}_t-\widehat{\bf x}_{t|t-1})({\bf x}_t-\widehat{\bf x}_{t|t-1})\T\}\label{cond_cov_a_priori}\\
\Sigma_{t|t}^{\bf x}&\triangleq\mathbb{E}\{({\bf x}_t-\widehat{\bf x}_{t|t})({\bf x}_t-\widehat{\bf x}_{t|t})\T\}.\label{cond_cov_a_posteriori}
\end{align}
Then, $\{(\widehat{\bf x}_{t|t-1}, \widehat{\bf x}_{t|t}, \Sigma_{t|t-1}^{\bf x}, \Sigma_{t|t}^{\bf x}): t\in\mathbb{N}_0^n\}$ are computed recursively forward in time as follows:
\begin{align}
\begin{split}
&\widehat{\bf x}_{t|t}=\widehat{\bf x}_{t|t-1}+{\bf k}^{\bf z}_t{\bf I}_t^{\bf z},~\widehat{\bf x}_{0|-1}=\bar{x}_0,\\
&\widehat{\bf x}_{t|t-1}=A_{t-1}\widehat{\bf x}_{t-1|t-1},\\
&\Sigma_{t|t-1}^{\bf x}={A}_{t-1}\Sigma^{\bf x}_{t-1|t-1}{A}_{t-1}\T+\Sigma_{{\bf w}_{t-1}},~\Sigma^{\bf x}_{0|-1}=\Sigma_{{\bf x}_0},\\
&{\bf I}_t^{\bf z}={\bf z}_t-\mathbb{E}\{{\bf z}_t|{\bf z}^{t-1}\}=C_t({\bf x}_t-\widehat{\bf x}_{t|t-1})+{\bf n}_t, \mbox{(Innovations)}\\
&\Sigma_{{\bf I}^{\bf z}_t}=C_t\Sigma^{\bf x}_{t|t-1}C_t\T+\Sigma_{{\bf n}_t}\\
&{\bf k}^{\bf z}_t=\Sigma^{\bf x}_{t|t-1}C_t\T\Sigma_{{\bf I}^{\bf z}_t}^{-1} ~\mbox{(Kalman ~Gain)},\\
&\Sigma_{t|t}^{\bf x}=\Sigma^{\bf x}_{t|t-1}-\Sigma^{\bf x}_{t|t-1}C_t\T\Sigma_{{\bf I}^{\bf z}_t}^{-1}C_t\Sigma^{\bf x}_{t|t-1},
\end{split}\label{KF_classical}
\end{align}
where ${\bf I}_t^{\bf z}$ is an orthogonal process independent of $(\widehat{\bf x}_{t|t-1},{\bf x}^{t-1}, {\bf z}^{t-1},{\bf y}^{t-1})$ and $\Sigma_{t|t}^{\bf x}\succeq{0}$ and $\Sigma^{\bf x}_{t|t-1}\succ{0}$.
\end{lemma}
\begin{IEEEproof}
The proof follows using every standard textbook on state estimation and filtering theories, see e.g., \cite{anderson:1979,kailath:2000,simon:2006,vanschuppen:2021} thus we omit it.
\end{IEEEproof}
Before we proceed to the next result, we state as a corollary the special case of time-varying fully observable multivariate Gauss-Markov process.
\begin{corollary}\label{corollary:onlemma_classical_kf}(Special case of Lemma \ref{lemma:classical_kf}) Suppose that the system model in \eqref{state_process_tv}, \eqref{observation_process_tv} is simplified to time-varying fully observable multivariate Gauss-Markov process, i.e.,
\begin{align}
{\bf x}_{t+1}&=A_t{\bf x}_{t}+{\bf w}_t,~{\bf x}_0=\bar{x},\label{state_process_1_tv}\\
{\bf z}_{t}&={\bf x}_{t},~t\in\mathbb{N}_0^n.\label{full_observation_process_tv}
\end{align}
Then, the KF recursions of \eqref{KF_classical} simplify as follows: $\widehat{\bf x}_{t|t}={\bf x}_t$, $\widehat{\bf x}_{t|t-1}=A_{t-1}{\bf x}_{t-1}$, $\Sigma_{t|t-1}^{\bf x}=\Sigma_{{\bf I}^{\bf z}_t}=\Sigma_{{\bf w}_{t-1}}$, $\Sigma_{t|t}^{\bf x}=0$ and ${\bf k}_t^{\bf z}=I_p$.
\end{corollary}
\begin{IEEEproof}
The derivation is straightforward using properties of conditional expectation and the fact that $C_t=I_p,~\forall{t}$, and $\Sigma_{{\bf n}_t}=0,~\forall{t}$.
\end{IEEEproof}
}
{
Before we prove a main structural result, we prove an adaptation of a result derived for random variables in \cite[p. 35]{cover-thomas:2006} to causal processes.
\begin{proposition}\label{prop:dpi}(Data processing inequality) Consider the joint process $\{({\bf x}_t,{\bf z}_t,{\bf y}_t):~t\in\mathbb{N}_0^n\}$. For each $t=0,1,\ldots,n$, let the statistic ${\bm \xi}_t=f({\bf z}^t)$. Then, 
\begin{align}
\sum_{t=0}^nI({\bm \xi}^t;{\bf y}_t|{\bf y}^{t-1})\leq\sum_{t=0}^n{I}({\bf z}^t;{\bf y}_{t}|{\bf y}^{t-1}),\label{inequality}
\end{align}
for any $n$, assuming $I({\bm \xi}^t;{\bf y}_t|{\bf y}^{t-1})<\infty$, ${I}({\bf z}^t;{\bf y}_{t}|{\bf y}^{t-1})<\infty, \forall{t}$. Moreover, the inequality in \eqref{inequality} holds with equality if 
\begin{align}
{\bf P}(dy_t|y^{t-1},\xi^t,z^t)={\bf P}(dy_t|y^{t-1},\xi^t), ~\forall{t\in\mathbb{N}_0^n}.\label{cond_indep_sufficient}
\end{align}
In that case, ${\bm \xi}^t$ is called {\it sufficient statistic} of the process ${\bf y}_t$ parametrized by ${\cal Y}^{t-1}$ because it contains all the information of ${\bf z}^t$ about ${\bf y}_t$ parametrized by ${\cal Y}^{t-1}$ at each instant of time.
\end{proposition}
\begin{IEEEproof}
The lower bound in \eqref{inequality} follows because the conditional independence ${\bf P}(dy_t|y^{t-1},\xi^t,z^t)={\bf P}(dy_t|y^{t-1},z^t), ~\forall{t=0,1,\ldots,n}$ is always true. Clearly, if \eqref{cond_indep_sufficient} holds the inequality in \eqref{inequality} holds with equality. 
\end{IEEEproof}
}
\par {Using Proposition \ref{prop:dpi} we prove structural sufficient conditions to ensure that \eqref{inequality} holds with equality for jointly Gaussian multivariate processes.
\begin{lemma}(Structural sufficient conditions for equality of \eqref{inequality})\label{lemma:suff_stat_gaussian}
Suppose that $\{({\bf x}_t,{\bf z}_t,{\bf y}_t):~t\in\mathbb{N}_0^n\}$ is a jointly Gaussian multivariate process. Moreover, let ${\bm \xi}_t=\mathbb{E}\{{\bf x}_t|{\bf z}^t\}$ (the MMSE estimator of ${\bf x}_t$ given ${\bf z}^t$). Then, \eqref{inequality} holds with equality if $C_t\in\mathbb{R}^{m\times{p}}$ in \eqref{observation_process_tv} is full row rank at each $t$. 
\end{lemma}
\begin{IEEEproof}
The proof is based on the structural properties of the optimal minimizer of the general problem in \eqref{finite_time_remote_nrdf_modified}. Observe that the following hold
\begin{align}
{\bf P}(dy_t|y^{t-1},z^t)&\stackrel{(a)}={\bf P}(dy_t|y^{t-1},z^t,{\xi}^t)\nonumber\\
&\stackrel{(b)}={\bf P}(dy_t|y^{t-1},{I^{\bf z}}^t,{\xi}^t)\nonumber\\
&\stackrel{(c)}={\bf P}(dy_t|y^{t-1},{\xi}^t),\label{equality_jointly_gaussian}
\end{align}
where $(a)$ follows because ${\bm \xi}_t=\mathbb{E}\{{\bf x}_t|{\bf z}^{t}\}$ and is consistent with the conditional independence ${\bf P}(dy_t|y^{t-1},\xi^t,z^t)={\bf P}(dy_t|y^{t-1},z^t), ~\forall{t=0,1,\ldots,n}$ of Proposition \ref{prop:dpi}; $(b)$ follows from Lemma \ref{lemma:classical_kf} because from the innovations process we have ${\bf z}_t={\bf I}^{\bf z}_t+\mathbb{E}\{{\bf z}_t|{\bf z}^{t-1}\}=C_tA_{t-1}{\bm \xi}_{t-1}+{\bf I}_t^{\bf z}$; $(c)$ follows because from Lemma \ref{lemma:classical_kf} we have ${\bm \xi}_t=A_{t-1}{\bm \xi}_{t-1}+{\bf k}_t{\bf I}_t^{\bf z}$ that for $\{0,1,\ldots,t\}$ includes the whole information about ${{\bf I}^{\bf z}}^t$ if ${\bf I}_t^{\bf z}={\bf k}_t^\dagger({\bm \xi}_t-A_{t-1}{\bm \xi}_{t-1})$, where ${\bf k}_t^\dagger=({\bf k}_t\T{\bf k}_t)^{-1}{\bf k}_t\T$ is a pseudo-inverse matrix with full column rank, which is the case if $C_t\in\mathbb{R}^{m\times{p}}$ in \eqref{observation_process_tv} is full row rank at each $t$. Hence, following Proposition \ref{prop:dpi} we proved that the conditional independence constraint \eqref{cond_indep_sufficient} holds and this in turn implies that for jointly Gaussian processes, ${\bm \xi}^t$ is a sufficient statistic about ${\bf z}^t$ for the process ${\bf y}_t$ parametrized by ${\cal Y}^{t-1}$. This completes the proof.
\end{IEEEproof}
\begin{remark}\label{remark:dim_red_phen}(Connection to \cite{tanaka:2015}) It should be emphasized that a similar result with Lemma \ref{lemma:suff_stat_gaussian}  was obtained in \cite[Lemma 2]{tanaka:2015} by taking {\it \'a priori} the structure of matrix $C_t$ in \eqref{observation_process_tv} and claiming that the pre-KF in his cascade realization is causally invertible thus an information lossless operation. In our case, we follow a structural approach reminiscent of the one proposed in \cite[p. 20, Eq. (II.135)-(II.138)]{charalambous2020new} by showing equality of the optimal minimizers via a sufficient statistic approach. Clearly, if the matrix $C_t\in\mathbb{R}^{m\times{p}}$ in \eqref{observation_process_tv} is not full row rank, then, the inequality in \eqref{inequality} is strict.
\end{remark}
}
\par { Next, we study the structure of the amended distortion constraint in the convex optimization problem of \eqref{finite_time_remote_nrdf_modified} obtained for jointly Gaussian processes. Specifically,
\begin{align}
\hat{d}(z^t,y_t)=&\mathbb{E}_{{\bf x}_t|{\bf z}^t={z}^t}\{||{\bf x}_t-{\bf y}_t||^2\}\nonumber\\
\stackrel{{\bm \xi}_t=\mathbb{E}\{{\bf x}_t|{\bf z}^t\}}=&\mathbb{E}_{{\bf x}_t|{\bf z}^t={z}^t}\{||{\bf x}_t-{\bm \xi}_t+{\bm \xi}_t-{\bf y}_t||^2\}\nonumber\\
\stackrel{(i)}=&\mathbb{E}_{{\bf x}_t|{\bf z}^t={z}^t}\{||{\bf x}_t-{\bm \xi}_t||^2\}+||{\bm \xi}_t-{\bf y}_t||^2\nonumber\\
\stackrel{(ii)}=&\trace(\Sigma_{t|t}^{\bf x})+||{\bm \xi}_t-{\bf y}_t||^2\label{modified cost}
\end{align}
where $(i)$ follows because for jointly Gaussian processes ${\bm \xi}_t$ is the optimal MMSE estimator of  ${\bf x}_t$ given ${\bf z}^t$ and from the orthogonality principle; $(ii)$ follows by definition of the {\it \'a posteriori} error covariance of the optimal MMSE obtained from the KF recursions in Lemma \ref{lemma:classical_kf}. Finally, the amended distortion that corresponds to the distortion constraint in \eqref{finite_time_remote_nrdf_modified} is obtained by taking the expectation with respect to the joint distribution of $\{({\bf z}_t,{\bf y}_t):~t\in\mathbb{N}_0^n\}$ in \eqref{modified cost} and then the summation which will give 
\begin{align}
\sum_{t=0}^n\trace(\Sigma_{t|t}^{\bf x})+\sum_{t=0}^n\mathbb{E}\{||{\bm \xi}_t-{\bf y}_t||^2\}.\label{distortion_jointly Gaussian}
\end{align}
Putting all the pieces together, we can reformulate \eqref{finite_time_remote_nrdf_modified} (and its asymptotic limit) to a convex problem for jointly Gaussian processes as follows.
\begin{definition}\label{def:remote_nrdf_gaussian}(Indirect NRDF for partially observable Gaussian processes) Suppose that the process $\{({\bf x}_t, {\bf z}_t, {\bf y}_t):~t\in\mathbb{N}_0^n\}$ is jointly Gaussian and matrix $C_t\in\mathbb{R}^{m\times{p}}$ in \eqref{observation_process_tv} is full row rank. Then, the indirect or remote NRDF \eqref{finite_time_remote_nrdf} and its asymptotic limit (provided it exists) can be reformulated as follows
\begin{align}
&R_{[0,n],\ind}^{G}(D-D^{\min}_{[0,n]})\nonumber\\
&=\inf_{\substack{{\bf P}(dy_t|y^{t-1},{\xi}^t):t\in\mathbb{N}_0^n \\ \frac{1}{n+1}\sum_{t=0}^n\mathbb{E}\{||{\bm \xi}_t-{\bf y}_t||^2\}\leq{D-D_{[0,n]}^{\min}}}}\sum_{t=0}^nI({\bm \xi}^t;{\bf y}_t|{\bf y}^{t-1}),\label{finite_time_remote_nrdf_gaussian}\\
&R_{\ind}^{G}(D-D_{[0,\infty]}^{\min})\nonumber\\
&=\inf_{\substack{{\bf P}(dy_t|y^{t-1},{\xi}^t):t=0,1,\ldots,\infty \\ \limsup_{n\longrightarrow\infty}\frac{1}{n+1}\sum_{t=0}^n\mathbb{E}\{||{\bm \xi}_t-{\bf y}_t||^2\}\leq{D-D_{[0,\infty]}^{\min}}}}\bar{R},\label{infinite_time_remote_nrdf_gaussian}
\end{align} 
where in \eqref{finite_time_remote_nrdf_gaussian} $(D-D_{[0,n]}^{\min})\in[0,\infty]$, $D_{[0,n]}^{\min}=\frac{1}{n+1}\sum_{t=0}^n\trace(\Sigma_{t|t}^{\bf x})$, in \eqref{infinite_time_remote_nrdf_gaussian} $D_{[0,\infty]}^{\min}=\limsup_{n\longrightarrow\infty}D_{[0,n]}^{\min}$ and $\bar{R}\triangleq\limsup_{n\longrightarrow\infty}\frac{1}{n+1}\sum_{t=0}^n{I}({\bm \xi}^t;{\bf y}_t|{\bf y}^{t-1})$.
\end{definition}
Next, we stress a few technical comments on Definition \ref{def:remote_nrdf_gaussian}.
\begin{remark}(On Definition \ref{def:remote_nrdf_gaussian}) {\bf (1)} The information measure \eqref{finite_time_remote_nrdf_gaussian} has a finite solution if we ensure that $D-D_{[0,n]}^{\min}\in(0,\infty]$ with $D_{[0,n]}^{\min}<\infty$; {\bf (2)} One can take the more stringent {\it pointwise} MSE distortion constraint in \eqref{finite_time_remote_nrdf_gaussian} in which case the problem in finite time horizon becomes 
\begin{align}
&R_{[0,n],\ind}^{G}(\{D_t-D^{\min}_{t}\}_{t=0}^n)\nonumber\\
&=\inf_{\substack{{\bf P}(dy_t|y^{t-1},{\xi}^t):~t\in\mathbb{N}_0^n \\ \mathbb{E}\{||{\bm \xi}_t-{\bf y}_t||^2\}\leq{D_t-D_{t}^{\min}},~\forall{t}}}\sum_{t=0}^nI({\bm \xi}^t;{\bf y}_t|{\bf y}^{t-1}),\label{finite_time_remote_nrdf_gaussian_pointwise_dist}
\end{align}
where $D_t^{\min}=\trace(\Sigma^{\bf x}_{t|t}),$ and $D_t-D_t^{\min}\in(0,\infty]$ with $D_t^{\min}<\infty$, for any $t$. {\bf (3)} The lower bound \eqref{finite_time_remote_nrdf_gaussian} shows an interesting resemblance to the classical remote RDF obtained for $\IID$ memoryless Gaussian processes or random variables using non-causal coding \cite{wolf-ziv:1970}. In particular, similar to that case, the distortion constraint in \eqref{finite_time_remote_nrdf_gaussian} consists of two parts of which only one affects the rates. As a result the other part can be essentially subtracted from the given distortion level. {\bf (4)} Definition \ref{def:remote_nrdf_gaussian} is not the same as the definition obtained in \cite[Eq. (16)]{tanaka:2015}. Therein the author assumes a lower bound with a {\it soft pointwise} MSE distortion constraint whereas we consider a lower bound subject to hard average total distortion constraints. {\bf (5)} To the best of the authors' knowledge, \cite{kostina:2019a} has never proved the information measure in Definition \ref{def:remote_nrdf_gaussian} for jointly Gaussian processes either in finite time or in the asymptotic limit.
\end{remark}
}
\par {The following is a key structural property on our objective function in \eqref{finite_time_remote_nrdf_gaussian} for the development of our results. 
\begin{proposition}\label{proposition:structural_result}(Structural properties of \eqref{finite_time_remote_nrdf_gaussian}) If the process $\{{\bm \xi}_t:~t\in\mathbb{N}_0^n\}$ admits the Markov realization obtained from the KF recursions of \eqref{KF_classical} under the structural property on matrix $C_t$ of the observations process \eqref{observation_process_tv}, then,  \eqref{finite_time_remote_nrdf_gaussian} simplifies to the following
\begin{align}
&R_{[0,n],\ind}^{G}(D-D^{\min}_{[0,n]})\nonumber\\
&=\inf_{\substack{{\bf P}(dy_t|y^{t-1},{\xi}_t):t\in\mathbb{N}_0^n \\ \frac{1}{n+1}\sum_{t=0}^n\mathbb{E}\{||{\bm \xi}_t-{\bf y}_t||^2\}\leq{D-D_{[0,n]}^{\min}}}}\sum_{t=0}^nI({\bm \xi}_t;{\bf y}_t|{\bf y}^{t-1}).\label{finite_time_remote_nrdf_gaussian_markov}
\end{align}
\end{proposition}
\begin{IEEEproof} Since the process $\{{\bm \xi}_t:~t\in\mathbb{N}_0^n\}$ admits the Markov realization in the KF recursions in \eqref{KF_classical}, i.e., ${\bm \xi}_t=A_{t-1}{\bm \xi}_{t-1}+{\bf k}_t^{\bf z}{\bf I}_t^{\bf z}$, then, the convex optimization problem in \eqref{finite_time_remote_nrdf_gaussian} is having implicit recursions similar to the ones described in \cite[Theorem 4.1]{stavrou:2018siam} obtained via dynamic programming backward in time \cite{bertsekas:2005}. The only difference is that the source distribution is simply replaced by the distribution $\{{\bf P}(d{\xi}_t|{\xi}_{t-1}):~t\in\mathbb{N}_0^n\}$. This completes the derivation.
\end{IEEEproof}
Using Proposition \ref{proposition:structural_result}, we will shortly provide {\it for the first time}, the {\it complete characterization in finite time} of the indirect NRDF for  time-varying partially observable  multidimensional jointly Gaussian processes. }

\par {As a first step, we need the following helpful lemma which is a non-trivial generalization of the classical KF algorithm and a generalization of a similar result obtained in \cite{stavrou:2018siam}.
\begin{lemma}(Realization of $\{{\bf P}^*(dy_t|y^{t-1},{\xi}_t):~t\in\mathbb{N}_0^n\}$)
\label{lemma:conditionally_gaussian_suff_stat}
For the system model in \eqref{state_process_tv}, \eqref{observation_process_tv}, suppose that the joint process $\{({\bf x}_t,~{\bf y}_t,~{\bf z}_t):~t\in\mathbb{N}^n_0\}$ is jointly Gaussian with $C_t\in\mathbb{R}^{m\times{p}}$ in \eqref{observation_process_tv} to be full row rank. Then, the following statements hold.\\
{\bf (1)} Any $\{{\bf P}^*(dy_t|y^{t-1},\xi_t):~t\in\mathbb{N}^n_0\}$ is realized by  
\begin{align}
{\bf y}_t=& H_t\left({\bm \xi}_t -\widehat{{\bm \xi}}_{t|t-1}\right)+ \widehat{\bm \xi}_{t|t-1}+ {\bf v}_t, \ \ t\in\mathbb{N}^n_0, \label{conditionally_gaussian_suff_stat}
\end{align}
where $\widehat{{\bm \xi}}_{t|t-1}\triangleq\mathbb{E}\{{\bm \xi}_t|{\bf y}^{t-1}\}$,~$\{{\bf v}_t\in\mathbb{R}^p\sim{\cal N}(0; \Sigma_{{\bf v}_t}): ~t\in\mathbb{N}^n_0\}$ is an independent Gaussian process independent of $\{({\bf w}_t,~{\bf n}_t): ~t \in {\mathbb N}_0^{n}\}$ and ${\bf x}_0$, and  $\{H_t\in\mathbb{R}^{p\times{p}}:~~t\in\mathbb{N}^n_0\}$ are time-varying deterministic matrices (to be designed). \\
Moreover, the innovations process $\{{\bf I}^{\bm \xi}_t\in\mathbb{R}^p:~t\in\mathbb{N}^n_0\}$ of \eqref{conditionally_gaussian_suff_stat} is the orthogonal process given by
\begin{align}
{\bf I}^{\bm \xi}_t={\bf y}_t-\mathbb{E}\{{\bf y}_t|{\bf y}^{t-1}\}=H_t \left({\bm \xi}_t -\widehat{{\bm \xi}}_{t|t-1} \right)+{\bf v}_{t}, \label{innovations_process_suff_stat}
\end{align}
where  ${\bf I}^{\bm \xi}_t\sim{\cal N}(0;\Sigma_{{\bf I}^{\bm \xi}_t})$,~$\Sigma_{{\bf I}^{\bm \xi}_t}=H_t\Sigma^{\bm \xi}_{t|t-1}{H}_t\T+\Sigma_{{\bf v}_t}$ with~$\Sigma^{\bm \xi}_{t|t-1}\triangleq\mathbb{E}\left\{({\bm \xi}_t-\widehat{\bm \xi}_{t|t-1})({\bm \xi}_t-\widehat{\bm \xi}_{t|t-1})\T\right\}$. \\
{\bf (2)} Let $\widehat{\bm \xi}_{t|t}\triangleq\mathbb{E}\{{\bm \xi}_t|{\bf y}^{t}\},~\Sigma_{t|t}^{\bm \xi}\triangleq\mathbb{E}\{({\bm \xi}_t-\widehat{\bm \xi}_{t|t})({\bm \xi}_t-\widehat{\bm \xi}_{t|t})\T\}$. Then, $\{(\widehat{\bm \xi}_{t|t-1},~\Sigma^{\bm \xi}_{t|t-1}, \widehat{\bm \xi}_{t|t},~\Sigma^{\bm \xi}_{t|t}):  t\in\mathbb{N}^n_0\}$ satisfy the following generalized discrete-time forward KF recursions:
\begin{align}
\begin{split}
&\widehat{\bm \xi}_{t|t}=\widehat{\bm \xi}_{t|t-1}+{\bf k}^{\bm \xi}_t{\bf I}^{\bm \xi}_t,\\
&\widehat{\bm \xi}_{t|t-1}=A_{t-1}\widehat{\bm \xi}_{t-1|t-1},~\widehat{\bm \xi}_{0|-1}=\bar{\bm \xi}_0\\
&\Sigma^{\bm \xi}_{t|t-1}={A}_{t-1}\Sigma^{\bm \xi}_{t-1|t-1}{A}_{t-1}\T+{\bf k}^{\bf z}_t\Sigma_{{\bf I}^{\bf z}_t}{{\bf k}^{\bf z}_t}\T,~\Sigma_{0|-1}^{\bm \xi}=\Sigma_{{\bf k}^{\bf z}_0\Sigma_{{\bf I}^{\bf z}_0}{{\bf k}^{\bf z}_0}\T},\\
&{\bf k}^{\bm \xi}_t=\Sigma^{\bm \xi}_{t|t-1}{H}_t\T\Sigma_{{\bf I}^{\bm \xi}_t}^{-1} ~\mbox{(Kalman ~Gain)},\\
&\Sigma_{t|t}^{\bm \xi}=\Sigma_{t|t-1}^{\bm \xi}-\Sigma_{t|t-1}^{\bm \xi}{H}_t\T\Sigma_{{\bf I}^{\bm \xi}_t}^{-1}H_t\Sigma^{\bm \xi}_{t|t-1},
\end{split}\label{KF_1_suff_stat}
\end{align}
where $\Sigma_{t|t}^{\bm \xi}\succeq{0}$ and $\Sigma_{t|t-1}^{\bm \xi}\succeq{0}$ (because ${\bf k}^{\bf z}_t\Sigma_{{\bf I}^{\bf z}_t}{{\bf k}^{\bf z}_t}\T\succeq{0}$).\\
{\bf (3)} The characterization of ${R}^{G}_{[0,n],\ind}(D-D_{[0,n],\ind}^{\min}) $ that achieves \eqref{conditionally_gaussian_suff_stat} is given by
\begin{align}
&{R}^{G}_{[0,n],\ind}(D-D_{[0,n]}^{\min})= \nonumber\\
&\inf_{\substack{H_t\in\mathbb{R}^{p\times{p}},~\Sigma_{{\bf v}_t}\succeq{0}\\ \Sigma^{\bm \xi}_{t|t}\succeq{0},~\Sigma^{\bm \xi}_{t|t-1}\succeq{0}\\
\frac{1}{n+1}\sum_{t=0}^n\trace\left({\cal G}\right)\leq{D-D_{[0,n]}^{\min}}}} \frac{1}{2}\sum_{t=0}^n\left[\log\frac{|\Sigma^{\bm \xi}_{t|t-1}|}{|\Sigma^{\bm \xi}_{t|t}|}\right]^{+},
\label{initial:optimization_suff_stat}
\end{align} 
where 
\begin{align}
{\cal G}=(I_p-H_t)\Sigma^{\bm \xi}_{t|t-1}(I_p-H_t)\T+\Sigma_{{\bf v}_t},\nonumber
\end{align} 
for some $D-D_{[0,n]}^{\min}\in  [0, \infty]$.
\end{lemma}
\begin{IEEEproof}
{\bf (1)} Since the joint process $\{({\bf x}_t, {\bf z}_t, {\bf y}_t):~t\in\mathbb{N}_0^n\}$ is assumed to be jointly Gaussian, then, $\{{\bf P}^*(dy_t|y^{t-1},\xi_t):~t\in\mathbb{N}_0^n\}$ is conditionally Gaussian, and we can obtain the orthogonal realization 
\begin{align}
&{\bf y}_t=H_t{\bm \xi}_t+R_t({\bf y}^{t-1})+{\bf v}_t,~t\in\mathbb{N}_0^n, \label{conditionally_gaussian_reproduction_suff_stat}
\end{align}
where $R_t({\bf y}^{t-1})\triangleq{\Gamma}_{t-1}{\bf y}^{t-1}$,~${\bf P}^*(\cdot|y^{t-1},\xi_t)\sim {\mathcal{N}}(H_t{\bm \xi}_t+\Gamma_{t-1}{\bf y}^{t-1}; \Sigma_{{\bf v}_t})$, with $\{(H_t, \Gamma_{t-1}):~t\in\mathbb{N}_0^n\}$ being deterministic matrices of appropriate dimensions. For such  realization, $I({\bm \xi}_t; {\bf y}_t|{\bf y}^{t-1})$ does not depend on $R_t(\cdot),~\forall t\in\mathbb{N}_0^n$. Moreover,
\begin{align}
\mathbb{E}\left\{||{\bm \xi}_t-{\bf y}_t||^2\right\}&=\mathbb{E}\left\{||(I_p-{H}_t){\bm \xi}_t-R_t({\bf y}^{t-1})||^2\right\} \nonumber\\
&+\trace\left(\Sigma_{{\bf v}_t}\right)\nonumber\\
&\stackrel{(\star)}\geq\mathbb{E}\left\{||(I_p-{H}_t){\bm \xi}_t-R^*_t({\bf y}^{t-1})||^2\right\}\nonumber\\
& +\trace\left(\Sigma_{{\bf v}_t}\right),\nonumber
\end{align}
where $(\star)$ holds with equality if $R_t(\cdot)=R^*_t(\cdot)= (I_p-{H}_t) \widehat{\bm \xi}_{t|t-1},~\forall t\in\mathbb{N}_0^n$.  {\bf (2)} This follows from the discrete-time KF equations. {\bf (3)} The characterization that achieves \eqref{conditionally_gaussian_suff_stat} is obtained from {\bf (1)}, {\bf (2)} and the definition of conditional mutual information $I({\bm \xi}_t;{\bf y}_t|{\bf y}^{t-1})$ at each time instant $t$.  This completes the proof. 
\end{IEEEproof}
}
{At this point we need to stress some important technical comments on Lemma \ref{lemma:conditionally_gaussian_suff_stat} which is essentially an intermediate step towards the complete characterization of the problem in finite time (i.e., it is not involved in the final optimization problem).}
{
\begin{remark}\label{remark:rank_def_innovations}(On Lemma \ref{lemma:conditionally_gaussian_suff_stat}) {\bf (1)} If in the KF recursions of Lemma \ref{lemma:conditionally_gaussian_suff_stat} we have $\Sigma_{{\bf I}^{\bm \xi}_t}\succeq{0}$, i.e., $\rank(\Sigma_{{\bf I}^{\bm \xi}_t})=l<p$, then, we can use a dimension reduction approach as follows. First, we need to find the pseudo-inverse matrix of $\Sigma_{{\bf I}_t}^{\bm \xi}$ using singular value decomposition (SVD) (or eigendecomposition) \cite{barnett:1990}. For example, recall that the eigendecomposition of $\Sigma_{{\bf I}_t}^{\bm \xi}=U\Sigma{U}\T$ where the orthogonal matrix ${U}=\begin{bmatrix}
U_1 &U_2
\end{bmatrix}
$ 
with $U_1\in\mathbb{R}^{p\times{l}}$, $U_2\in\mathbb{R}^{p\times{(p-l)}}$ and $\Sigma=\diag\big(\mu_{\Sigma_{{\bf I}_t}^{\bm \xi},1},\ldots,\mu_{\Sigma_{{\bf I}_t}^{\bm \xi},l},~\underbrace{0,\ldots,0}_{(p-l)~elements}\big)$. Now the pseudo-inverse matrix can be simply constructed by $\Sigma^{\dagger}_{{\bf I}^{\bm \xi}_t}={U}\Sigma^{+}U\T$ where $\Sigma^{+}=\diag\big(\mu^{-1}_{\Sigma_{{\bf I}_t}^{\bm \xi},1},\ldots,\mu^{-1}_{\Sigma_{{\bf I}_t}^{\bm \xi},l},\underbrace{0,\ldots,0}_{(p-l)~elements}\big)$  is formed by taking the inverse of all non-zero elements of $\Sigma$. The next step is to exclude the degenerated rows and columns (i.e., the $p-l$-dimensional null space) from $(U, \Sigma^{+})$ hence obtaining $\Sigma^{\dagger}_{{\bf I}^{\bm \xi}_t}={U}_l{\Sigma}_l^{+}{U}_l\T$ with ${U}_l\in\mathbb{R}^{l\times{l}}$ and ${\Sigma}_l^{+}=\diag\big(\mu^{-1}_{\Sigma_{{\bf I}_t}^{\bm \xi},1},\ldots,\mu^{-1}_{\Sigma_{{\bf I}_t}^{\bm \xi},l})$. In other words, we can obtain for some $t$ the pseudo-inverse matrix $\Sigma^{\dagger}_{{\bf I}^{\bm \xi}_t}$, i.e., $\Sigma^{\dagger}_{{\bf I}^{\bm \xi}_t}=U_l\Sigma^{+}_lU_l\T$ where $U_l$ and $\Sigma^{+}_l$ are matrices with $(p-l)$ degenerated rows and columns deleted. In the end, the reproduction process admits a dimension reduction of $p-l$ elements, that is,  $\{{\bf y}_t\in\mathbb{R}^l:~t\in\mathbb{N}_0^n\}$; {\bf (2)} The characterization in \eqref{initial:optimization_suff_stat} is general and at this point we did not give conditions to ensure existence of a finite solution. Such conditions are pivotal and will be provided in the sequel; {\bf (3)} Similar to Corollary \ref{corollary:onlemma_classical_kf}, one can recover from Lemma \ref{lemma:conditionally_gaussian_suff_stat} the special case of time-varying fully observable multivariate Gauss-Markov processes. In particular, for the system model in \eqref{state_process_1_tv}, \eqref{full_observation_process_tv}, we obtain using Corollary \ref{corollary:onlemma_classical_kf} that in Lemma \ref{lemma:conditionally_gaussian_suff_stat} ${\bm \xi}_t={\bf x}_t,~\forall{t}$, $\{{\bf P}^*(dy_t|y^{t-1},\xi_t)\equiv{\bf P}^*(dy_t|y^{t-1},x_t):~t\in\mathbb{N}^n_0\}$ and $\widehat{{\bm \xi}}_{t|t-1}=\mathbb{E}\{{\bf x}_t|{\bf y}^{t-1}\}$. The analysis will give precisely the result first derived in \cite[Lemma 5.2]{stavrou:2018siam}. 
\end{remark}
}
\par {The next theorem gives the complete finite dimensional characterization of \eqref{finite_time_remote_nrdf_gaussian_markov} for partially observable Gaussian processes when the end-to-end system is jointly Gaussian subject to an average total $\mse$ distortion constraint. It also reveals the optimal linear Gaussian test-channel distribution (forward test-channel realization) that corresponds to this problem. 
\begin{theorem}(Complete characterization of \eqref{finite_time_remote_nrdf_gaussian_markov} for jointly Gaussian processes)\label{theorem:complete_characterization_suff_stat}
The information measure in \eqref{finite_time_remote_nrdf_gaussian_markov} corresponds to the following characterization
\begin{align}
&{R}^{G}_{[0,n],\ind}(D-D_{[0,n]}^{\min})= \nonumber\\
&\inf_{\substack{0\preceq{\Sigma^{\bm \xi}_{t|t}}\preceq\Sigma^{\bm \xi}_{t|t-1}\\ 
\frac{1}{n+1}\sum_{t=0}^n\trace\left(\Sigma^{\bm \xi}_{t|t}\right)\leq{D-D_{[0,n]}^{\min}}}} \frac{1}{2}\sum_{t=0}^n\left[\log\frac{|\Sigma^{\bm \xi}_{t|t-1}|}{|\Sigma^{\bm \xi}_{t|t}|}\right]^{+},\label{complete_characterization_suff_stat}
\end{align}
for some $D-D_{[0,n]}^{\min}\in[D_{\min},D_{\max}]\subset[0, D_{\max}]$. Moreover, if $\Sigma_{{\bf I}_t^{\bm \xi}}\succ{0}$ the above characterization, is achieved by a linear Gaussian ``test channel'' ${\bf P}^*(dy_t|y_{t-1},\xi_t)$ of the form  
\begin{align}
{\bf y}_t={H}_t{\bm \xi}_t+(I_p-{H}_t)A{\bf y}_{t-1}+{\bf v}_t,~t\in\mathbb{N}_0^n,\label{optimal_realization_suff_stat_1}
\end{align}
where ${\bf y}_t\in\mathbb{R}^p$, with ${\bf y}_{-1}=0$, and
\begin{align}
\begin{split}
&H_t\Sigma_{t|t-1}^{\bm \xi}\triangleq\Sigma_{t|t-1}^{\bm \xi}-\Sigma_{t|t}^{\bm \xi}\succeq{0},~\Sigma_{{\bf v}_t}\triangleq\Sigma_{t|t}^{\bm \xi}H_t\T\succeq{0}.
\end{split}\label{scalings_suff_stat_1}
\end{align} 
Otherwise, if $\Sigma_{{\bf I}_t^{\bm \xi}}\succeq{0}$ with $\rank(\Sigma_{{\bf I}_t^{\bm \xi}})=l<p$, the characterization in \eqref{complete_characterization_suff_stat} corresponds to a linear Gaussian ``test channel'' ${\bf P}^*(dy_t|y_{t-1},\xi_t)$ of the form \eqref{optimal_realization_suff_stat_1} with reduced dimension ${\bf y}_t\in\mathbb{R}^l$ ($l<p$) such that the scaling \eqref{scalings_suff_stat_1} hold for $l<p$. 
\end{theorem}
\begin{IEEEproof} 
From MSE estimation theory we know that the $\mse$ inequality $\sum_{t=0}^n\mathbb{E}\left\{||{\bm \xi}_t-{\bf y}_t||^2\right\}\geq \sum_{t=0}^n\mathbb{E}\left\{||{\bm \xi}_t-\widehat{\bm \xi}_{t|t}||^2\right\}$ holds for all $({H}_t,~\Sigma_{{\bf v}_t})$, $t\in\mathbb{N}_0^n$, and it is achieved if and only if $\widehat{\bm \xi}_{t|t}={\bf y}_t$. Sufficient conditions for the latter to hold are (i) $\widehat{\bm \xi}_{t|t-1}\equiv\mathbb{E}\{{\bf y}_t|{\bf y}^{t-1}\}$ and (ii) ${\bf k}^{\bm \xi}_t=I_p$. Note that (i) holds by the general KF algorithm in Lemma \ref{lemma:conditionally_gaussian_suff_stat}. The choice of \eqref{scalings_suff_stat_1} satisfies (ii) as long as $\Sigma_{{\bf I}_t^{\bm \xi}}\succ{0}$ hence a smaller distortion for a given rate is achieved and also the Markov realization in \eqref{optimal_realization_suff_stat_1} holds. If $\Sigma_{{\bf I}_t^{\bm \xi}}\succeq{0}$, then, following Remark \ref{remark:rank_def_innovations}, we simply seek for the pseudo-inverse matrix $\Sigma^{\dagger}_{{\bf I}^{\bm \xi}_t}=U_{l}\Sigma^{+}_{l}U_l\T$ and exclude from the system the degenerated rows and columns of dimension $p-l$. 
Under this dimension reduction, the sufficient conditions  (i), (ii) hold and the test-channel realization with the appropriate scalings follows. 
\end{IEEEproof}
\begin{remark}\label{remark:suff_cond_characterization}(Sufficient conditions for existence of solution in \eqref{complete_characterization_suff_stat}) Sufficient conditions for existence of an optimal solution with finite value in \eqref{complete_characterization_suff_stat} are as follows:
\begin{itemize}
\item[{\bf (C1)}] $D_{[0,n]}^{\min}<\infty$ for any finite $n$;
\item[{\bf (C2)}] $D-D_{[0,n]}^{\min}>0$ (non-zero distortion) which implies the strict linear matrix inequality (LMI) constraint $0\prec\Sigma_{t|t}^{\bm \xi}\preceq\Sigma_{t|t-1}^{\bm \xi},~\forall{t}$.  
\end{itemize}
\end{remark}
From Theorem \ref{theorem:complete_characterization_suff_stat} it can be easily checked that via Corollary \ref{corollary:onlemma_classical_kf} we can recover the complete characterization in finite time of the NRDF for time-varying fully observable Gauss-Markov processes with MSE distortion first derived in \cite{stavrou:2018siam}.}

{Next, we give the optimal numerical solution of the characterization of Theorem \ref{theorem:complete_characterization_suff_stat} as long as the conditions of Remark \ref{remark:suff_cond_characterization} hold. 
\begin{theorem}\label{theorem:numer_col_finite_time}(Optimal numerical solution of \eqref{complete_characterization_suff_stat}) Compute forward in time via \eqref{KF_classical} $\{(\Sigma_{t|t}^{\bf x}, \Sigma_{t|t-1}^{\bf x}):~t\in\mathbb{N}_0^n\}$ making sure that the conditions of Remark \ref{remark:suff_cond_characterization} hold. Then, the solution of \eqref{complete_characterization_suff_stat} is semidefinite representable as follows:
\begin{itemize}
\item[{\bf (1)}] Suppose $A_t\in\mathbb{R}^{p\times{p}}$ is full rank and ${\bf k}_{t+1}^{\bf z}\Sigma_{{\bf I}_{t+1}^{\bf z}}{\bf k}_{t+1}^{{\bf z}\T}\succeq{0}$. Moreover, introduce the decision variable $\Gamma^{1}_t\succ{0}$ and let the factorization of the singular matrix ${\bf k}_{t+1}^{\bf z}\Sigma_{{\bf I}_{t+1}^{\bf z}}{\bf k}_{t+1}^{{\bf z}\T}\triangleq{B}_{t+1}B_{t+1}\T$. Then, for $D>D_{[0,n]}^{\min}$ we obtain
\begin{align}
&{R}^{G}_{[0,n],\ind}(D-D_{[0,n]}^{\min})=\nonumber \\
&\min_{\substack{\{\Sigma_{t|t}^{\bm \xi}\succ{0}, \Gamma^{1}_t\succ{0}\}_{t=0}^n\\{0\prec\Sigma^{\bm \xi}_{0|0}}\preceq\Sigma^{\bm \xi}_{0|-1}\\
 0\prec\Sigma^{\bm \xi}_{t+1|t+1}\preceq\Sigma^{\bm \xi}_{t+1|t},~t\in\mathbb{N}_0^{n-1}\\
 \Sigma_{n|n}^{\bm \xi}=\Gamma^{1}_{n}}}\frac{1}{2}\sum_{t=0}^n\log|\Gamma_t^{1}|^{-1}+c_1,\label{numer_sol_suff_stat:finite_time_1}\\ 
&s.~t.~\qquad\frac{1}{n+1}\sum_{t=0}^n\trace\left(\Sigma^{\bm \xi}_{t|t}\right)\leq{D-{D}_{[0,n]}^{\min}}\nonumber\\ 
&\qquad\qquad\left[ \begin{array}{cc}
I_p-\Gamma^{1}_t & B_{t+1}\T \nonumber\\
B_{t+1} & \Sigma_{t+1|t}^{\bm \xi}\end{array}\right]\succeq 0,~t\in\mathbb{N}_0^{n-1}  \nonumber
\end{align}
\end{itemize} 
where $c_1=\frac{1}{2}\log|\Sigma_{0|-1}^{\bm \xi}|+\sum_{t=0}^{n-1}\log\abs\left(|A_t|\right)$.
\begin{itemize}
\item[{\bf (2)}] Suppose ${\bf k}_{t+1}^{\bf z}\Sigma_{{\bf I}_{t+1}^{\bf z}}{{\bf k}_{t+1}^{\bf z}}\T\succ{0},~\forall{t}$, and introduce the decision variable $\Gamma^{2}_t\succ{0}$. Then, for $D>D_{[0,n]}^{\min}$ we obtain
\begin{align}
&{R}^{G}_{[0,n],\ind}(D-D_{[0,n]}^{\min})= \nonumber\\
&\min_{\substack{\{\Sigma_{t|t}^{\bm \xi}\succ{0}, \Gamma^{2}_t\succ{0}\}_{t=0}^n\\{0\prec\Sigma^{\bm \xi}_{0|0}}\preceq\Sigma^{\bm \xi}_{0|-1}\\
 0\prec\Sigma^{\bm \xi}_{t+1|t+1}\preceq\Sigma^{\bm \xi}_{t+1|t},~t\in\mathbb{N}_0^{n-1}\\
 \Sigma_{n|n}^{\bm \xi}=\Gamma^{2}_{n}}}\frac{1}{2}\sum_{t=0}^n\log|\Gamma_t^{2}|^{-1}+c_2,\label{numer_sol_suff_stat:finite_time_2}\\ 
&s.~t.~\qquad\frac{1}{n+1}\sum_{t=0}^n\trace\left(\Sigma^{\bm \xi}_{t|t}\right)\leq{D-{D}_{[0,n]}^{\min}}\nonumber\\ 
&\qquad\qquad\left[ \begin{array}{cc}
\Sigma_{t|t}^{\bm \xi}-\Gamma_t^2 & \Sigma_{t|t}^{\bm \xi}A_t\T \nonumber\\
A_t\Sigma_{t|t}^{\bm \xi} & \Sigma_{t+1|t}^{\bm \xi} \end{array}\right]\succeq 0,~t\in\mathbb{N}_0^{n-1}  \nonumber
\end{align}
where $c_2=\frac{1}{2}\log|\Sigma_{0|-1}^{\bm \xi}|+\frac{1}{2}\sum_{t=0}^{n-1}\log|{\bf k}_{t+1}^{\bf z}\Sigma_{{\bf I}_{t+1}^{\bf z}}{{\bf k}_{t+1}^{\bf z}}|$.
\end{itemize}
\end{theorem}
\begin{IEEEproof}
See Appendix~\ref{proof:theorem:numerical_solution:finite_time}.
\end{IEEEproof}
}
{
Next, we stress some technical comments on Theorem \ref{theorem:numer_col_finite_time}.
\begin{remark}\label{remark:ontheorem_numer_sol_finite_time}(On Theorem \ref{theorem:numer_col_finite_time}) {\bf (1)} To compute the optimal numerical solutions in Theorem \ref{theorem:numer_col_finite_time} is computationally very expensive. First we need to compute $\{(\Sigma_{t|t-1}^{\bf x}, \Sigma_{t|t}^{\bf x}):~t\in\mathbb{N}_0^n\}$ of Lemma \ref{lemma:classical_kf} both of dimension $p\times{p}$, which correspond to approximately ${\cal O}(p^{2.376})$ operations for each time instant $t$, then, to engage SDP algorithm of which the most computationally expensive step is the Cholesky factorization that requires, in general, approximately ${\cal O}(p^3)$ operations at each time instant $t$. Some additional analysis on the arithmetic complexity of the SDP algorithm is provided in \cite[Sec. IV-C]{tanaka:2017}. In fact as we demonstrate in the sequel (see Table \ref{table:comparison}) even for the single stage case at high dimensional problems, the SDP algorithm (with the steady-state covariance matrices obtained by the pre-KF recursions) operates extremely slow. Hence finding alternative optimal or near-optimal algorithmic approaches with reasonable computational complexity aligned with the state of the art large scale networks that operate using computationally limited resources remains an intriguing open problem. {\bf (2)} Theorem \ref{theorem:numer_col_finite_time} continues to hold with appropriate changes if we consider the stronger pointwise distortion constraint, i.e., $\trace\left(\Sigma^{\bm \xi}_{t|t}\right)\leq{D_t-{D}_{t}^{\min}},~D_t>D_t^{\min},~D_t^{\min}<\infty,~\forall{t}$. 
\end{remark}
\paragraph*{Closed-form solutions for scalar processes} To gain further insights on the solution of the problem for time-varying systems in finite time, in what follows, we propose a solution (via a dynamic reverse waterfilling algorithm) under average total MSE distortion constraints and a closed-form solution under pointwise MSE distortion. Consider the scalar-valued system model of \eqref{state_process_tv}, \eqref{observation_process_tv} of the form
\begin{align}
\begin{split}
{\bf x}_{t+1}&=\alpha_t{\bf x}_{t}+{\bf w}_t,~{\bf x}_0=\bar{x},\\
{\bf z}_t&=c_t{\bf x}_t+{\bf n}_t,~~t\in\mathbb{N}_0^n,
\end{split}\label{scalar_noisy_observ_system}
\end{align}
where $\alpha_t\in\mathbb{R}$ and $c_t\in\mathbb{R}\setminus\{0\}$ are non-random, ${\bf x}_0\in\mathbb{R}\sim{\cal N}(0;\sigma^2_{{\bf x}_0})$ is the initial state, ${\bf w}_t\in\mathbb{R}\sim{\cal N}(0;\sigma^2_{{\bf w}_t})$, $\sigma^2_{{\bf w}_t}>0$ is an independent sequence,  ${\bf n}_t\in\mathbb{R}\sim{\cal N}(0;\sigma^2_{{\bf n}_t})$, $\sigma^2_{{\bf n}_t}\geq{0}$, is an independent sequence, independent of $\{{\bf w}_t:~t\in\mathbb{N}_0^n\}$, whereas ${\bf x}_0$ is independent of $\{({\bf w}_t,{\bf n}_t):~t\in\mathbb{N}_0^n\}$. Before we proceed, we denote $\Sigma_{t|t}^{\bf x}\equiv\sigma_{{\bf x}_{t|t}}^2$, $\Sigma_{t|t-1}^{\bf x}\equiv\sigma_{{\bf x}_{t|t-1}}^2$, $\Sigma_{t|t}^{\bm \xi}\equiv\sigma_{{\bm \xi}_{t|t}}^2$, $\Sigma_{t|t-1}^{\bm \xi}\equiv\sigma_{{\bm \xi}_{t|t-1}}^2$, ${\bf k}^{\bf z}_{t}\Sigma_{{\bf I}^{\bf z}_t}{{\bf k}^{\bf z}_{t}}\T=\frac{c_t^2\sigma^4_{{\bf x}_{t|t-1}}}{c_t^2\sigma^2_{{\bf x}_{t|t-1}}+\sigma^2_{{\bf n}_t}}\equiv\sigma^2_{{\upsilon}_t},~$for any $t\in\mathbb{N}_0^n$. Additionally, we rewrite the general characterization of Theorem \ref{theorem:complete_characterization_suff_stat} for scalar processes under the assumption that the total rates yield a finite solution, i.e.,
\begin{align}
&{R}^{G}_{[0,n],\ind}(D-D_{[0,n]}^{\min})= \nonumber\\
&\inf_{\substack{0<\sigma^2_{{\bm \xi}_{t|t}}\leq\sigma^2_{{\bm \xi}_{t|t-1}}\\ 
\frac{1}{n+1}\sum_{t=0}^n\left(\sigma^2_{{\bm \xi}_{t|t}}\right)\leq{D-D_{[0,n]}^{\min}}}} \frac{1}{2}\sum_{t=0}^n\log\left(\frac{\sigma^2_{{\bm \xi}_{t|t-1}}}{\sigma^2_{{\bm \xi}_{t|t}}}\right),\label{complete_characterization_suff_stat_scalar}
\end{align}
where $D-D_{[0,n]}^{\min}>0$, $D_{[0,n]}^{\min}=\frac{1}{n+1}\sum_{t=0}^n\sigma^2_{{\bf x}_{t|t}}<\infty$.
In the next theorem, we give the optimal solution of \eqref{complete_characterization_suff_stat_scalar} via a dynamic reverse-waterfilling algorithm.
\begin{theorem}\label{theorem:opti_num_sol_scalar_tv}(Optimal solution of \eqref{complete_characterization_suff_stat_scalar}) The optimal parametric solution of \eqref{complete_characterization_suff_stat_scalar} can be computed as follows:
\begin{align}
&{R}^{G}_{[0,n],\ind}(D-D_{[0,n]}^{\min})=\frac{1}{2}\sum_{t=0}^n\log\left(\frac{\sigma^2_{{\bm \xi}_{t|t-1}}}{\sigma^2_{{\bm \xi}_{t|t}}}\right),\label{par_sol_scalar_tv}
\end{align}
such that $\sigma^2_{{\bm \xi}_{t|t}}>0$ is computed at each time instant as follows:
\begin{align}
\sigma^2_{{\bm \xi}_{t|t}}=\begin{cases} \sigma^{2,*}_{{\bm \xi}_{t|t}}~&\mbox{if}~\sigma^{2,*}_{{\bm \xi}_{t|t}}<\sigma^2_{{\bm \xi}_{t|t-1}}\\
\sigma^2_{{\bm \xi}_{t|t-1}}~&\mbox{if}~\sigma^{2,*}_{{\bm \xi}_{t|t}}\geq\sigma^2_{{\bm \xi}_{t|t-1}}
\end{cases},~\forall{t}, \label{rev_water_scalar_tv}
\end{align}
with $\sum_{t=0}^n\sigma^2_{{\bm \xi}_{t|t}}=(n+1)(D-D_{[0,n]}^{\min})$ and
\begin{align}
\sigma^{2,*}_{{\bm \xi}_{t|t}}=\begin{cases}
\frac{1}{\beta_{t,t+1}}\left(\sqrt{1+\frac{\beta_{t,t+1}}{\theta^*}}-1\right),~\forall{t}\in\mathbb{N}_0^{n-1}\\
\frac{1}{2\theta^*},~t=n,
\end{cases},\label{sol_kkt_tv}
\end{align}
where $\theta^*>0$, $\beta_{t,t+1}\triangleq\frac{2\alpha_t^2}{\sigma_{\upsilon_{t+1}}^2}$, and $D>D_{[0,n]}^{\min}$ with $D_{[0,n]}^{\min}<\infty$.
\end{theorem}
\begin{IEEEproof}
See Appendix~\ref{proof:theorem:opti_num_sol_scalar_tv}. 
\end{IEEEproof} 
\begin{remark}\label{remark:ontheorem_scalar_rev_water_tv}(On Theorem \ref{theorem:opti_num_sol_scalar_tv}) Suppose that in \eqref{scalar_noisy_observ_system} we set $c_t=1$ and ${\bf n}_t=0,~\forall{t}$. Then, using Corollary \ref{corollary:onlemma_classical_kf} it can be easily observed that $\sigma^2_{\upsilon_t}=\sigma^2_{{\bf w}_{t-1}}$, ${\beta}_{t,t+1}=\frac{2\alpha_t^2}{\sigma_{{\bf w}_t}^2}$ and $D_{[0,n]}^{\min}=0,~\forall{t}$, and we recover \cite[Theorem 1]{stavrou:2021tac2}.
\end{remark}
In what follows, we propose an algorithmic embodiment of the optimal solution of Theorem \ref{theorem:opti_num_sol_scalar_tv}.
\begin{varalgorithm}{1}
\caption{Implementation of Theorem \ref{theorem:opti_num_sol_scalar_tv}}
{
\begin{algorithmic}
\STATE {\textbf{Initialize:} number of time-steps $n$; error tolerance $\epsilon$; nominal minimum and maximum value of $\theta$, denoted by $\theta^{\min}$ and $\theta^{\max}$; initial variance ${\sigma}^2_{{\bf x}_{0|-1}}=\sigma^2_{{\bf x}_{0}}$; set values for $\{(\alpha_t, \sigma^2_{{\bf w}_t}, c_t, \sigma^2_{{\bf n}_t}):~t\in\mathbb{N}_0^n\}$ of  \eqref{scalar_noisy_observ_system}.}
\FOR {$t=0:n$} 
\STATE {Compute $(\sigma^2_{{\bf x}_{t|t}}, \sigma^2_{{\bf x}_{t|t-1}})$ via \eqref{KF_classical}.}
\ENDFOR
\STATE{Compute $D_{[0,n]}^{\min}=\frac{1}{n+1}\sum_{t=0}^n\sigma^2_{{\bf x}_{t|t}}<\infty$; set the distortion level $D>D_{[0,n]}^{\min}$; Pick some $\theta\in[\theta^{\min}, \theta^{\max}]$; $\text{flag}=0$.}
\WHILE{$\text{flag}=0$}
\FOR {$t=0:n$}
\STATE {Compute $\sigma^{2,*}_{{\bm \xi}_{t|t}}$ according to \eqref{sol_kkt_tv}.}
\STATE {Compute $\sigma^{2}_{{\bm \xi}_{t|t}}$ according to \eqref{rev_water_scalar_tv}.}
\IF {$t<n$}
\STATE{Compute $\sigma^2_{{\bm \xi}_{t+1|t}}$ according to $\sigma^2_{{\bm \xi}_{t+1|t}}\triangleq\alpha^2_{t}\sigma^{2}_{{\bm \xi}_{t|t}}+\sigma^2_{{\upsilon}_{t+1}}$.}
\ENDIF
\ENDFOR
\IF {$\frac{1}{n+1}\sum_{t=0}^n{\sigma^2_{{\bm \xi}_{t|t}}}-(D-D_{[0,n]}^{\min})\geq\epsilon$}
\STATE {Set $\theta^{\min}={\theta}$.}
\ELSE
\STATE {Set $\theta^{\max}={\theta}$.}
\ENDIF
\IF {$\theta^{\max}-\theta^{\min}\geq\frac{\epsilon}{n+1}$}
\STATE{Compute $\theta=\frac{(\theta^{\min}+\theta^{\max})}{2}$.}
\ELSE
\STATE {$\text{flag}\leftarrow 1$}
\ENDIF
\ENDWHILE
\STATE {\textbf{Output:} $\{\sigma^2_{{\bm \xi}_{t|t}}:~t\in\mathbb{N}_0^n\}$, $\{\sigma^2_{{\bm \xi}_{t|t-1}}:~t\in\mathbb{N}_0^n\}$, for a given distortion level $D-D_{[0,n]}^{\min}$.}
\end{algorithmic}
\label{algo1}
}
\end{varalgorithm}
\begin{remark}\label{remark:complexity:algo1}(On Algorithm \ref{algo1})
Algorithm \ref{algo1} ensures linear convergence in finite time via a {\it bisection method} for a given error tolerance $\epsilon$ by picking as starting points appropriate nominal range of values for $\theta$ (i.e., $\theta^{\min}$ and $\theta^{\max}$). The convergence of bisection method implies that $\theta$ converges, hence ~$\frac{1}{n+1}\sum_{t=0}^n{\sigma^2_{{\bm \xi}_{t|t}}}\longrightarrow(D-D_{[0,n]}^{\min})$ within the error tolerance $\epsilon$. We note that the nominal values of $\theta^{\min}$ and $\theta^{\max}$  vary depending on the data of the system model \eqref{scalar_noisy_observ_system}. The most computationally expensive operation in Algorithm \ref{algo1} is the {\it for loop} and the {\it bisection method} that yield a time complexity of approximately ${\cal O}(n\log(n))$ (linearithmic time complexity). In Fig. \ref{fig:average_time_complexity} we illustrate a numerical simulation of the average running time needed for Algorithm \ref{algo1} to execute (vs) the time horizon $n$ when the error tolerance is $\epsilon=10^{-9}$. For this simulation we consider that each $n$ is the mean of $10000$ time instants.
{\bf (2)} As Remark \ref{remark:ontheorem_scalar_rev_water_tv} suggests, for time-varying fully observable Gauss-Markov processes Algorithm \ref{algo1} recovers the algorithm proposed in \cite[Algorithm 1]{stavrou:2021tac2}.
\end{remark}
\begin{figure}[ht]
\centering
\includegraphics[width=\columnwidth]{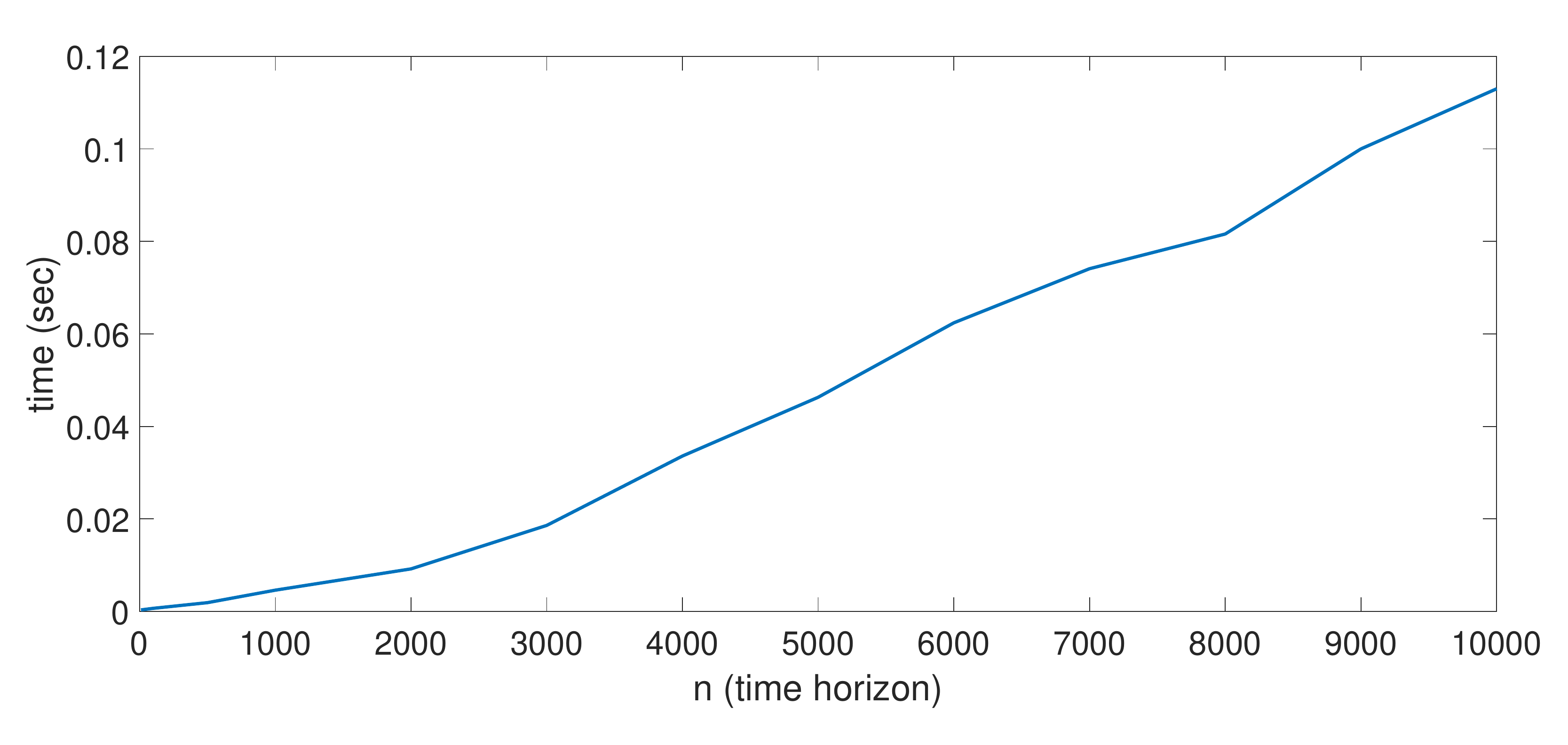}
\caption{ {Illustration of the average running time needed
for Algorithm \ref{algo1} to execute for
$10000$ instances. Simulations were performed in MATLAB
R2017b and tested on a single CPU with an Intel Core i7
processor at 2.6 GHz, 16 GB RAM and Windows 10.}}\label{fig:average_time_complexity}
\end{figure}
In the next corollary, we give for the first time, a closed form expression of the system in \eqref{scalar_noisy_observ_system} under pointwise MSE distortion constraints.
\begin{corollary}\label{corollary:closed_form_scalar_tv}(An optimal closed form solution) Find forward in time $\{(\sigma^2_{{\bf x}_{t|t}}, \sigma^2_{{\bf x}_{t|t-1}}):~t\in\mathbb{N}_0^n\}$ via \eqref{KF_classical} and let $D_t>D_t^{\min}=\sigma^2_{{\bf x}_{t|t}},~\forall{t}$. Then,  the closed form solution of \eqref{complete_characterization_suff_stat_scalar} under a pointwise MSE distortion constraint is given as follows
\begin{align}
&{R}^{G}_{[0,n],\ind}(\{D_t-D_{t}^{\min}\}_{t=0}^n)=\frac{1}{2}\sum_{t=0}^n\left[\log\left(\frac{\sigma_{{\bm \xi}_{t|t-1}}^2}{D_t-D_t^{\min}}\right)\right]^{+},
\end{align}\label{scalar_closed_form_pointwise_mse_tv}
where $\sigma_{{\bm \xi}_{t|t-1}}^2=\alpha_{t-1}^2(D_{t-1}-D_{t-1}^{\min})+\sigma_{{\upsilon}_t}^2$.
\end{corollary}
\begin{IEEEproof}
The proof is similar to Theorem \ref{theorem:numer_col_finite_time} by employing KKT conditions hence it is omitted. 
\end{IEEEproof}
}
\par {For the special case of time-varying fully-observable Gauss-Markov processes it can be easily seen following precisely Remark \ref{remark:ontheorem_scalar_rev_water_tv} that we can recover \cite[Corollary 2]{stavrou:2018lcss}.
}

%
%
{
\section{Complete characterization and optimal computational methods: infinite time horizon}\label{sec:complete_char_compute_infinite_time}
}
\par {In this section, we analyse the asymptotic limit of \eqref{finite_time_remote_nrdf_gaussian_markov}. 
To do it, we first restrict our system model \eqref{state_process_tv}, \eqref{observation_process_tv} to time-invariant processes, i.e., $A_t=A$, $\Sigma_{{\bf w}_t}=\Sigma_{\bf w}$, $C_t=C$, $\Sigma_{{\bf n}_t}=\Sigma_{\bf n},~\forall{t}$. Second, we apply known results for the convergence of the discrete time Riccati equation (DRE) of Lemma \ref{lemma:classical_kf}. These results can be found for instance in \cite[Chapter 7.3]{simon:2006}, \cite[Appendix E]{kailath:2000} or \cite{vanschuppen:2021}. Before we state a lemma, we note that in the sequel, we adopt for simplicity the following notation
\begin{align}
\begin{split}
\Sigma_t&=\Sigma_{t|t}^{\bf x},~\mbox{and}~\Sigma=\lim_{t\longrightarrow\infty}\Sigma_{t} \\
\Pi_t&=\Sigma_{t|t-1}^{\bf x},~\mbox{and}~\Pi=\lim_{t\longrightarrow\infty}\Pi_{t}\\
\bar{\Sigma}_t&={\bf k}_t^{\bf z}\Sigma_{{\bf I}_t^{\bf z}}{{\bf k}_t^{\bf z}}\T,~\mbox{and}~\bar{\Sigma}=\lim_{t\longrightarrow\infty}\bar{\Sigma}_t.
\end{split}
\end{align}\label{state_state_sol_RE}
\begin{lemma}\label{lemma:necessary_suff} \cite{kailath:2000},\cite{simon:2006} (Necessary and sufficient conditions for convergence of the time-invariant DRE of Lemma \ref{lemma:classical_kf} to a unique stabilizing solution) Let $(A, \Sigma_{\bf w}, C, \Sigma_{\bf n})\in\mathbb{R}^{p\times{p}}\times\mathbb{R}^{p\times{p}}\times\mathbb{R}^{m\times{p}}\times\mathbb{R}^{m\times{m}}$. Then, the DRE that corresponds to Lemma \ref{lemma:classical_kf} is the following
\begin{align}
\Pi_{t}=&A\Pi_{t-1}A\T-A\Pi_{t-1}C\T(C\Pi_{t-1}C\T+\Sigma_{\bf n})^{-1}C\Pi_{t-1}{A}\T\nonumber\\
&\qquad+\Sigma_{\bf w},~t\in\mathbb{N}_0,\label{dre}
\end{align}
where $\Pi_0\succ{0}$ (always positive definite). Moreover, the corresponding discrete time algebraic Riccati equation (DARE) is as follows
\begin{align}
\Pi=A\Pi{A}\T-A\Pi{C}\T(C\Pi{C}\T+\Sigma_{\bf n})^{-1}C{\Pi}A\T+\Sigma_{\bf w}.\label{dare}
\end{align}
Then, the following statement holds. Let the pair $(A, C)$ to be detectable and the pair $(A, \Sigma_{\bf w}^{\frac{1}{2}})$ to be stabilizable (or controllable on and outside the unit circle). Then, any solution of \eqref{dre}, i.e, $\{\Pi_t:~t\in\mathbb{N}_0\}$, is such that $\lim_{t\longrightarrow\infty}\Pi_t=\Pi$, $\Pi\succeq{0}$ for any $\Pi_0\succeq{0}$ which corresponds to the maximal unique stabilizing solution of \eqref{dare}. This further means that the steady-state KF, i.e., the limiting expression of $\widehat{\bf x}_{t|t}\equiv{\bm \xi}_t$ in \eqref{KF_classical} is asymptotically stable.
\end{lemma}
Next, we provide an example applied to scalar processes, to illustrate the concept of Lemma \ref{lemma:necessary_suff}.
\begin{example}\label{example:1}(Convergence of the time-invariant DRE for scalar processes) Consider the time-invariant version of the system model in \eqref{scalar_noisy_observ_system}, i.e., $\alpha_t=\alpha\in\mathbb{R},$ $\sigma_{{\bf w}_t}^2=\sigma_{{\bf w}}^2>0$, $c_t=c\in\mathbb{R}\setminus\{0\}$, $\sigma_{{\bf n}_t}^2=\sigma_{{\bf n}}^2\geq{0},~\forall{t}$. Then, the time-invariant scalar-valued DRE of \eqref{dre} is
\begin{align}
\begin{split}
\Pi_{t}=\alpha^2\Pi_{t-1}+\sigma_{\bf w}^2-\frac{\alpha^2c^2\Pi^2_{t-1}}{c^2\Pi_{t-1}+\sigma^2_{\bf n}},~t\in\mathbb{N}_0,
\end{split}\label{dre_scalar}
\end{align}
where $\Pi_0>{0}$. The corresponding scalar-valued DARE of \eqref{dare} is as follows
\begin{align}
\begin{split}
\Pi=\alpha^2\Pi+\sigma_{\bf w}^2-\frac{\alpha^2c^2\Pi^2}{c^2\Pi+\sigma^2_{\bf n}}.
\end{split}\label{dare_scalar}
\end{align}
Moreover, introduce the pairs $(a, c)$ and $(\alpha, (\sigma^2_{\bf w})^{\frac{1}{2}})$. Then, by definition, the pair $(\alpha, c)$ is always detectable and the pair $(\alpha, (\sigma^2_{\bf w})^{\frac{1}{2}})$ is always stabilizable (because by definition $\sigma_{\bf w}^2>0$). Hence, from Lemma \ref{lemma:necessary_suff} any solution of \eqref{dre_scalar} is such that $\lim_{t\longrightarrow\infty}\Pi_t=\Pi$, with $\Pi\geq{0}$ that corresponds to the unique stabilizing solution of \eqref{dare_scalar}. In what follows, we compute the closed form solution of $\Pi\geq{0}$. Note that \eqref{dare_scalar} can be reformulated to the quadratic equation $c^2\Pi^2+\gamma\Pi-\sigma^2_{\bf w}\sigma^2_{\bf n}=0,$ where $\gamma\triangleq((1-\alpha)^2\sigma^2_{\bf n}-c^2\sigma_{\bf w}^2)$, that gives the following two solutions
\begin{align}
\Pi=\begin{cases}
\frac{\sqrt{\gamma^2+4c^2\sigma_{\bf w}^2\sigma_{\bf n}^2}-\gamma}{2c^2} >0\\
\frac{-\sqrt{\gamma^2+4c^2\sigma_{\bf w}^2\sigma_{\bf n}^2}-\gamma}{2c^2}<0.
\end{cases}\label{closed_form_are_scalar}
\end{align}
Clearly, by conditions the negative solution of $\Pi$ is rejected. Hence from \eqref{closed_form_are_scalar} we have that the unique stabilizing solution is not only non-negative but also positive. This is because from by definition of our system model ($\Pi_0>0$ because $\sigma_{\bf w}^2>0$).
\paragraph*{\it Special cases} (i) Suppose that in \eqref{dre_scalar} we let $\sigma^2_{\bf n}=0$. Then, from \eqref{dare_scalar} we obtain $\Pi=\sigma_{\bf w}^2>0$. (ii) Suppose that in \eqref{dre_scalar} we let $\alpha=0$. Then, from \eqref{dare_scalar} we obtain $\Pi=\sigma_{\bf w}^2>0$. (iii) Suppose that in \eqref{dre_scalar}  we let $\alpha=0$, $\sigma_{\bf n}^2=0$. Then, from \eqref{dre_scalar} we obtain $\Pi=\sigma_{\bf w}^2>0$.
\end{example}
\begin{remark}(On Lemma \ref{lemma:necessary_suff}) In Lemma \ref{lemma:necessary_suff} we gave necessary and sufficient conditions for the DARE of the {\it \'a priori error covariance} in Lemma \ref{lemma:classical_kf} to converge to its steady state. We observed via Example \ref{example:1} that this value is always positive for scalar processes. Clearly, what we observe for scalar processes holds for multidimensional processes because $\Sigma_{\bf w}\succ{0}$ with $\Pi_t\succ{0},~\forall{t}$ (from Lemma \ref{lemma:classical_kf}). Moreover, the steady state of the {\it \'a priori error covariance} in Lemma \ref{lemma:classical_kf}  implies the convergence of the {\it \'a posteriori error covariance} as well. That case however is slightly different because we can allow an initial condition $\Sigma_0\succeq{0}$ (see Lemma \ref{lemma:classical_kf}). In fact as Corollary \ref{corollary:onlemma_classical_kf} suggests, if $\Sigma_{\bf n}=0$ and $C=I_p$, then $\Sigma_{t|t}^{\bf x}=0,~\forall{t}$, and as a result $\Sigma=\lim_{t\longrightarrow\infty}\Sigma_t=0$. 
\end{remark}
}
\par {Next, we prove a theorem where we give necessary and sufficient conditions for the pre-KF algorithm to converge to its steady-state values and conditions to ensure a time-invariant solution of the characterization in \eqref{complete_characterization_suff_stat}.
\begin{theorem}\label{theorem:existence_asymptotic_limit}(Asymptotic characterization of \eqref{complete_characterization_suff_stat}) Suppose that the system \eqref{state_process_tv}, \eqref{observation_process_tv} is restricted to time-invariant processes with the pair $(A, C)$ detectable and the pair $(A, \Sigma_{\bf w}^{\frac{1}{2}})$ stabilizable. Moreover restrict the test-channel distribution ${\bf P}(dy_t|y_{t-1},\xi_t)$ to be time-invariant and the output distribution ${\bf P}(dy_t|y_{t-1})$ to be time-invariant with a unique invariant distribution. Then, if $R_{[0,\infty],\ind}^{G}(D-D_{[0,\infty]}^{\min})<\infty$, for $D-D_{[0,\infty]}^{\min}\in(0,\infty]$, it is given by
\begin{align}
&{R}^{G}_{[0,\infty],\ind}(D-D_{[0,\infty]}^{\min})=\inf_{\substack{0\prec{\Sigma^{\bm \xi}}\preceq\Pi^{\bm \xi}\\ 
\trace\left(\Sigma^{\bm \xi}\right)\leq{D-D_{[0,\infty]}^{\min}}}} \frac{1}{2}\log\frac{|\Pi^{\bm \xi}|}{|\Sigma^{\bm \xi}|},\label{complete_characterization_suff_stat_infinite_horizon}
\end{align}
where $\Sigma^{\bm \xi}\succ{0}$ and $\Pi^{\bm \xi}\succ{0}$ are the time-invariant values of $\Sigma_{t|t}^{\bm \xi}$, and $\Sigma_{t|t-1}^{\bm \xi}$, respectively. Moreover, 
\begin{align}
\Pi^{\bm \xi}&=A\Sigma^{\bm \xi}A\T+\bar{\Sigma}\label{a_priori_tinv}\\
D_{[0,\infty]}^{\min}&=\trace(\Sigma).\label{steady_state_min_dist}
\end{align}
Finally, the above characterization, is achieved by a time-invariant linear Gaussian ``test channel'' ${\bf P}^*(dy_t|y_{t-1},\xi_t)$ of the form  
\begin{align}
{\bf y}_t={H}{\bm \xi}_t+(I_p-{H})A{\bf y}_{t-1}+{\bf v}_t,\label{optimal_realization_suff_stat_infinite_horizon}
\end{align}
where $H=I_p-\Sigma^{\bm \xi}(\Pi^{\bm \xi})^{-1}$ and ${\bf v}_t\sim{\cal N}(0;\Sigma_{\bf v}),~\Sigma_{\bf v}=\Sigma^{\bm \xi}H\T\succeq{0}$.
\end{theorem}
\begin{IEEEproof}
See Appendix \ref{proof:theorem:existence_asymptotic_limit}.
\end{IEEEproof}
\begin{remark}(On Theorem \ref{theorem:existence_asymptotic_limit}) {\bf (1)} From Theorem \ref{theorem:existence_asymptotic_limit} we can easily recover the known characterization in the infinite time horizon of the time-invariant fully observable Gauss-Markov processes (see, e.g., \cite[Theorem 3]{stavrou:2018}). In particular, suppose that the system model \eqref{state_process_1_tv}, \eqref{full_observation_process_tv} is restricted to time-invariant processes, i.e., $A_t=A$, $\Sigma_{{\bf w}_t}=\Sigma_{\bf w},~\forall{t}$. Then, from Corollary \ref{corollary:onlemma_classical_kf} we obtain ${\bf k}^{\bf z}\Sigma_{{\bf I}^{\bf z}}{{\bf k}^{\bf z}}\T=\Sigma_{\bf w}$ and $D_{[0,\infty]}^{\min}=0$ that when substituted in \eqref{complete_characterization_suff_stat_infinite_horizon}
recovers \cite[eq. (26)]{stavrou:2018}.
\end{remark}
In what follows, we give the optimal numerical solution of the problem in Theorem \ref{theorem:existence_asymptotic_limit}. 
\begin{corollary}\label{corollary:complete_optim_numer_infinite_horizon}(Optimal numerical solution of \eqref{complete_characterization_suff_stat_infinite_horizon})
The solution of the \eqref{complete_characterization_suff_stat_infinite_horizon} is semidefinite representable as follows:
\begin{itemize}
\item[(1)] Suppose $A\in\mathbb{R}^{p\times{p}}$ is full rank and ${\bf k}^{\bf z}\Sigma_{{\bf I}^{\bf z}}{{\bf k}^{\bf z}}\T\succeq{0}$. Moreover, introduce the decision variable $\Gamma^{1}\succ{0}$ and let the factorization of the singular matrix $\bar{\Sigma}\triangleq{B}B\T$. Then, for $D>D_{[0,\infty]}^{\min}$ we obtain
\begin{align}
&{R}^{G}_{\ind}(D-D_{[0,\infty]}^{\min})=\min_{\substack{ \Gamma^{1}\succ{0}\\ 0\prec\Sigma^{\bm \xi}\preceq\Pi^{\bm \xi}}}\frac{1}{2}\log|\Gamma^{1}|^{-1}+\log\abs\left(|A|\right),\label{numer_sol_suff_stat:infinite_time_1}\\ 
&s.~t.~\qquad\trace\left(\Sigma^{\bm \xi}\right)\leq{D-{D}_{[0,\infty]}^{\min}}\nonumber\\ 
&\qquad\qquad\left[ \begin{array}{cc}
I_p-\Gamma^{1} & B\T \nonumber\\
B & \Pi^{\bm \xi}\end{array}\right]\succeq 0.\nonumber
\end{align}
\item[(2)] Suppose $\bar{\Sigma}\succ{0}$. Moreover, introduce the decision variable $\Gamma^{2}\succ{0}$ with $\Sigma^{\bm \xi}\succ{0}$. Then, for $D>D_{[0,\infty]}^{\min}$ we obtain
\begin{align}
&{R}^{G}_{[0,\infty],\ind}(D-D_{[0,\infty]}^{\min})=\nonumber\\
&\min_{\substack{\Gamma^{2}_t\succ{0}\\
 0\prec\Sigma^{\bm \xi}\preceq\Pi^{\bm \xi}}}\frac{1}{2}\log|\Gamma^{2}|^{-1}+\frac{1}{2}\log|{\bf k}^{\bf z}\Sigma_{{\bf I}^{\bf z}}{{\bf k}^{\bf z}}|,\label{numer_sol_suff_stat:infinite_time_2}\\ 
&s.~t.~\qquad\trace\left(\Sigma^{\bm \xi}\right)\leq{D-{D}_{[0,\infty]}^{\min}}\nonumber\\ 
&\qquad\qquad\left[ \begin{array}{cc}
\Sigma^{\bm \xi}-\Gamma^2 & \Sigma^{\bm \xi}A\T \nonumber\\
A\Sigma^{\bm \xi} & \Pi^{\bm \xi} \end{array}\right]\succeq{0}.\nonumber
\end{align}
\end{itemize}
\end{corollary}
\begin{IEEEproof}
The proof is a special case of Theorem \ref{theorem:numer_col_finite_time} hence we omit it.
\end{IEEEproof}
}
\par {In what follows, we derive strong structural properties on  \eqref{complete_characterization_suff_stat_infinite_horizon} that allow for a simplified optimization problem which can be optimally solved via a reverse-waterfilling algorithm. 
\begin{proposition}\label{proposition:strong_structural_properties}(Strong structural properties on \eqref{complete_characterization_suff_stat_infinite_horizon}) Suppose that in the characterization of \eqref{complete_characterization_suff_stat_infinite_horizon} one of the following structures between $(A, \bar{\Sigma})$ hold.
\begin{itemize}
\item[(i)] Suppose that $A={\alpha}I_p$ (scalar matrix) and $\bar{\Sigma}\succeq{0}$;
\item[(ii)] Suppose that $A$ is real symmetric and $\bar{\Sigma}=\sigma^2_{\bar{\Sigma}}I_p$ (scalar matrix);
\item[(iii)] Suppose that $A=\bar{\Sigma}\succ{0}$;
\end{itemize}
Then $(A, \Sigma^{\bm \xi}, \bar{\Sigma})$ commute by pairs\footnote{{Details on this concept can be found in e.g., \cite[Theorem 21.13.1]{harville:1997}.}} and consequently $(\Sigma^{\bm \xi}, \Pi^{\bm \xi})$ commute. 
\end{proposition}
\begin{IEEEproof}
The derivation is done following similar arguments to \cite[Proposition 1]{stavrou:2021tac} thus we omit it.
\end{IEEEproof}
\begin{theorem}\label{theorem:rev_water}(Optimal numerical solution of \eqref{complete_characterization_suff_stat_infinite_horizon})   
Suppose that one of the conditions of Proposition \ref{proposition:strong_structural_properties} hold. Then, 
\begin{align}
\begin{split}
&{R}^{G}_{[0,\infty],\ind}(D-D_{[0,\infty]}^{\min})=\\
&\min_{\substack{0<\mu_{\Sigma^{\bm \xi},i}\leq\mu_{\Pi^{\bm \xi},i}\\ \sum_{i=1}^p\mu_{\Sigma^{\bm \xi},i}\leq{D}-D_{[0,\infty]}^{\min}}}\frac{1}{2}\sum_{i=1}^p\log\left(\frac{\mu_{\Pi^{\bm \xi},i}}{\mu_{\Sigma^{\bm \xi},i}}\right).
\end{split}\label{lower_bound_infinite_horizon}
\end{align}
Moreover, the optimal parametric solution of  \eqref{lower_bound_infinite_horizon} can be computed for $\mu_{\Sigma^{\bm \xi},i}>0$,~and any ${i}$ as follows:
\begin{align}
\mu_{\Sigma^{\bm \xi},i}=\begin{cases} \mu^*_{\Sigma^{\bm \xi},i}~&\mbox{if}~\mu^*_{\Sigma^{\bm \xi},i}<\mu_{\Pi^{\bm \xi},i}\\
\mu_{\Pi^{\bm \xi},i}~&\mbox{if}~\mu^*_{\Sigma^{\bm \xi},i}\geq\mu_{\Pi^{\bm \xi},i}
\end{cases},~\forall{i}, \label{rev_water_asympt_inf_hor_case2}
\end{align}
with $\sum_{i=1}^p\mu_{\Sigma^{\bm \xi},i}=(D-D_{[0,\infty]}^{\min})$ and
\begin{align}
\mu^*_{\Sigma^{\bm \xi},i}=\begin{cases}
\frac{1}{\mu_{\Upsilon,i}}\left(\sqrt{1+\frac{\mu_{\Upsilon,i}}{\theta^*}}-1\right),~\mbox{$\mu_{\Upsilon,i}>{0}$ for some $i$}\\
\frac{1}{2\theta^*},~\mbox{$\mu_{\Upsilon,i}={0}$}
\end{cases},\label{sol_kkt_tv_inf_hor_case2}
\end{align}
where $\theta^*>0$, $\mu_{\Upsilon,i}\triangleq\frac{2\mu_{A^2,i}}{\mu_{\bar{\Sigma},i}}>{0}$\footnote{{If $\bar{\Sigma}\succeq{0}$ we simply replace it $\bar{\Sigma}_\epsilon\succ{0}$ and consequently $\mu_{\bar{\Sigma},i}\equiv\mu_{\bar{\Sigma}_\epsilon,i}$.}} and $D>D_{[0,\infty]}^{\min}$ .
\end{theorem}
\begin{IEEEproof}
When $\bar{\Sigma}\succ{0}$ the derivation follows using similar steps of the derivation of \cite[Theorem 2]{stavrou:2021tac} and we omit it. However, if for instance in Proposition \ref{proposition:strong_structural_properties}, (i), $\bar{\Sigma}\succeq{0}$ we use a standard continuity argument, that is, there exists a $\delta>0$ such that $\bar{\Sigma}_\epsilon=\bar{\Sigma}+\epsilon{I}_p$ is nonsingular for all $\epsilon\in(0,\delta)$ (see, e.g., \cite[Theorem 2.9]{zhang:2011}). In other words, we create a $\bar{\Sigma}_\epsilon\succ{0}$, then following similar steps to the derivation of \cite[Theorem 2]{stavrou:2021tac} and taking in the computations that $\lim_{\epsilon\longrightarrow{0}^{+}}\bar{\Sigma}_{\epsilon}$ the result follows.
\end{IEEEproof}
An implementation of the reverse-waterfilling solution of Theorem \ref{theorem:rev_water} is given in Algorithm \ref{algo2}.
\begin{varalgorithm}{2}
\caption{{Implementation of Theorem \ref{theorem:rev_water}}}
{
\begin{algorithmic}
\STATE {\textbf{Initialize:} error tolerance $\epsilon$; nominal minimum and maximum value of $\theta$, i.e., $\theta^{\min}$ and $\theta^{\max}$; set values for $(A, \Sigma_{\bf w}, C, \Sigma_{\bf n})$ of  \eqref{scalar_noisy_observ_system} so that the par $(A, C)$ is detectable and the pair $(A, \Sigma_{\bf w}^{\frac{1}{2}})$ is stabilizable.}
\STATE{Find the unique stabilizing solution $\Pi$ and the steady-state value of $\Sigma$ via \eqref{dare} and compute  $D_{[0,\infty]}^{\min}=\trace(\Sigma)<\infty$; choose distortion level $D>\trace(\Sigma)$; Pick $\theta\in[\theta^{\min}, \theta^{\max}]$; find the eigenvalues of $(A, \bar{\Sigma})$, i.e., $\{\mu_{A,i}:~i=1\in\mathbb{N}_1^p\}$, $\{\mu_{\bar{\Sigma},i}:~i=1\in\mathbb{N}_1^p\}$ (in decreasing order); $\text{flag}=0$.}
\WHILE{$\text{flag}=0$}
\STATE {Compute $\mu_{\Sigma^{\bm \xi},i},~\forall{i},$ as follows:}
\FOR {$i=1:p$}
\STATE {Compute $\mu^*_{\Sigma^{\bm \xi},i}$ according to \eqref{sol_kkt_tv_inf_hor_case2}.}
\STATE {Compute $\mu_{\Sigma^{\bm \xi},i}$ according to \eqref{rev_water_asympt_inf_hor_case2}.}
\ENDFOR
\IF {$\theta^{\max}-\theta^{\min}\geq{\epsilon}$}
\STATE{Compute $\theta=\frac{(\theta^{\min}+\theta^{\max})}{2}$.}
\ELSE
\STATE {$\text{flag}\leftarrow 1$}
\ENDIF
\ENDWHILE
\STATE {\textbf{Output:} $\{\mu_{\Sigma^{\bm \xi},i}:~i\in\mathbb{N}_1^p\}$, $\{\mu_{\Pi^{\bm \xi},i}:~i\in\mathbb{N}_1^p\}$, for a given distortion level $D-\trace(\Sigma)$.}
\end{algorithmic}
\label{algo2}
}
\end{varalgorithm}
}
{
\begin{remark}\label{remark:compl_algo_2}(Complexity of Algorithm \ref{algo2})
Again the convergence of Algorithm \ref{algo2} is guaranteed for finite dimensional matrices due to the bisection method similar to Algorithm \ref{algo1}. The most computationally expensive parts in Algorithm \ref{algo2} are the {\it matrix multiplications in the computation of the DARE of the steady-state pre-KF recursions} which can have a time complexity of approximately ${\cal O}(p^3)$ followed by the {\it for loop} and the {\it bisection method} with approximately linearithmic time complexity similar to Algorithm \ref{algo1}, i.e., ${\cal O}(p\log(p))$. Hence the overall time complexity is approximately ${\cal O}(p^3+p\log(p))$. However, if we optimize matrix multiplication using for example the current state of the art computing approaches that allow time complexity of around ${\cal O}(p^{2.37286}))$ \cite{alman-williams:2021} the complexity can further reduce to ${\cal O}(p^{2.37286}+p\log(p))$. In Table \ref{table:comparison} we compare the general optimal solution obtained via SDP in Corollary \ref{corollary:complete_optim_numer_infinite_horizon} with the structural optimal solution obtained in Theorem \ref{theorem:rev_water} and implemented in Algorithm \ref{algo2} for the same input data and distortion level. For low dimensional vector systems (i.e., $p=10$) we compute the average computational time needed for $1000$ instances using both computational methods for an error tolerance of $\epsilon=10^{-9}$. We see that Algorithm \ref{algo2} is approximately $550$ times faster than SDP. For medium size vector systems (i.e., $p=100$) we perform the same experiment for $100$ instances with $\epsilon=10^{-7}$. The results show that Algorithm \ref{algo2} is approximately $17500$ times faster than SDP. We note that to obtain a result from SDP for $1000$ instances would require days therefore we did not attempt with the specific computer such experiment. In addition, it is likely that the result for both SDP and Algorithm \ref{algo2} would not change much. For high dimensional vector systems (i.e., $p=500$) the result is not-conclusive because SDP would take many days to give a relatively fair result even for $100$ instances. In contrast Algorithm \ref{algo2} operates fine as illustrated in Table \ref{table:comparison}. The results clearly demonstrate that Algorithm \ref{algo2} is much more appealing choice to use when solving problems with certain structure or systems with computationally limited resources as opposed to the SDP algorithm.
\renewcommand{\arraystretch}{1.2}
\begin{table}[!h]
\centering
    \begin{tabular}{  l | c | c }
    \hline
     \rowcolor[HTML]{a5ecb5}
    \textbf{Solver (Numb. dimens. $p=10$)} & \textbf{Mean} (sec) &\textbf{Numb. inst.} \\ \hline
    SDP (by default $\epsilon=10^{-9}$) & 0.7134 & 1000\\ \hline 
    Algorithm~\ref{algo2} ($\epsilon=10^{-9}$)&0.0013 & 1000  \\
    \hline
    \rowcolor[HTML]{b5ead1}
    \textbf{Solver (Numb. dimens. $p=100$)} & \textbf{Mean} (sec) & \textbf{Numb. inst.} \\ \hline
    SDP (by default $\epsilon=10^{-7}$) & 725.0770 & 100\\ \hline 
   Algorithm~\ref{algo2} ($\epsilon=10^{-7}$)&0.0412 & 100 \\   \hline
\rowcolor[HTML]{c2ebe9}
    \textbf{Solver (Numb. dimens. $p=500$)} & \textbf{Mean} (sec) &\textbf{Numb. inst.} \\ \hline
 SDP  & non-conclusive  & insufficient\\ \hline 
    Algorithm~\ref{algo2} ($\epsilon=10^{-9}$)&6.8997 & 1000\\   \hline     
\end{tabular}
\caption{{ Comparison of the computational time needed between \text{SDP} in Corollary \ref{corollary:complete_optim_numer_infinite_horizon} and Algorithm \ref{algo2}. Simulations were performed in MATLAB R2017b and tested on a single CPU with an Intel Core i7 processor at 2.6 GHz, 16 GB RAM and Windows 10.}}\label{table:comparison}
\end{table}
\renewcommand{\arraystretch}{1}
\end{remark}
}
 {We conclude this section, by finding the optimal analytical solution for the time-invariant version of the system model \eqref{scalar_noisy_observ_system}.
\begin{corollary}\label{corollary:scalar_time_inv}(Closed form solution: time-invariant scalar processes) Consider the characterization of Theorem \ref{theorem:existence_asymptotic_limit} restricted to time-invariant scalar Gaussian processes. Then for $D>D_{[0,\infty]}^{\min}=\Sigma$, the closed form solution of $R_{\ind}^G(D-\Sigma)$ is as follows
\begin{align}
R_{\ind}^G(D-\Sigma)=\frac{1}{2}\log\left(\alpha^2+\frac{\bar{\Sigma}}{D-\Sigma}\right)\label{closed_form_scalar_inf_hor}
\end{align}
where
\begin{align}
\bar{\Sigma}=\frac{c^2\Pi^2}{c^2\Pi+\sigma^2_{\bf n}},\label{ss_noise_scalar}
\end{align}
with $\Pi>0$ given by the unique stabilizing solution of \eqref{dare_scalar} whereas $\Sigma\geq{0}$ is given by the non-negative solution of the quadratic equation
\begin{align}
\alpha^2c^2\Sigma^2+\bar{\gamma}\Sigma-\sigma_{\bf w}^2\sigma_{\bf n}^2=0,\label{scalar_dare_posterior}
\end{align} 
where $\bar{\gamma}=(1-\alpha^2)\sigma_{\bf n}^2+c^2\sigma_{\bf w}^2$. 
\end{corollary}
\begin{IEEEproof}
For scalar processes, the characterization in Theorem \ref{theorem:existence_asymptotic_limit} simplifies to 
\begin{align}
R_{\ind}^G(D-D_{[0,\infty]}^{\min})=\min_{\substack{0<\Sigma^{\bm \xi}\leq\Pi^{\bm \xi}\\ \Sigma^{\bm \xi}\leq{D-D_{[0,\infty]}^{\min}}}}\frac{1}{2}\log\left(\frac{\Pi^{\bm \xi}}{\Sigma^{\bm \xi}}\right).\label{charact_inf_hor_scalar}
\end{align}
where $\Pi^{\bm \xi}=\alpha^2\Sigma^{\bm \xi}+\bar{\Sigma}$, $\bar{\Sigma}$ is given by \eqref{ss_noise_scalar} and $D_{[0,\infty]}^{\min}=\Sigma\geq{0}$, i.e., the unique stabilizing solution obtained for scalar processes given by \eqref{scalar_dare_posterior}. The problem in \eqref{charact_inf_hor_scalar} is convex with respect to $\Sigma^{\bm \xi}$ and the optimal solution follows by employing KKT conditions similar to Theorems \ref{theorem:opti_num_sol_scalar_tv}, and  \ref{theorem:rev_water}. It easy to see that the solution ensures $\Sigma^{\bm \xi}=D-D_{[0,\infty]}^{\min}=D-\Sigma$. Substituting the latter in $\Pi^{\bm \xi}$ and then substituting both $\Sigma^{\bm \xi}$ and $\Pi^{\bm \xi}$ in \eqref{charact_inf_hor_scalar} we obtain \eqref{closed_form_scalar_inf_hor} and the result follows.
\end{IEEEproof}
\paragraph*{\it Equivalent expressions and special cases for scalar processes} (i) We note that our closed form expression  \eqref{closed_form_scalar_inf_hor} coincides with the closed-form solution obtained via \cite[Corollary 1, Theorem 9]{kostina:2019a} (see also \cite[eq. (103)]{kostina:2020}) because the steady-state counterpart of the \'a posteriori error variance equation \eqref{KF_classical} implies the equality $\bar{\Sigma}=\Pi-\Sigma>0$;  (ii) Consider in Corollary \ref{corollary:scalar_time_inv} $c=1$, $\sigma_{\bf n}^2=0$. Then, using Example \ref{example:1} we obtain from \eqref{dare_scalar} that $\Pi=\sigma^2_{\bf w}>0$, from \eqref{scalar_dare_posterior} the steady state solution is $\Sigma=0$ and from \eqref{ss_noise_scalar} $\bar{\Sigma}=\sigma^2_{\bf w}>0$. By substituting these in \eqref{closed_form_scalar_inf_hor} we recover the known result obtained for time-invariant or stationary fully observable Gauss-Markov processes, see e.g., \cite[eq. (14)]{tatikonda:2004}, \cite[eq. (1.43)]{gorbunov:1972b}. 
}
%
%
{
\section{Numerical Simulations}\label{sec:num_sim}
}
\par {In this section, we provide two examples with numerical simulations for some of the major results of this paper.
\begin{example}(Optimal numerical solutions and comparison with \cite{kostina:2019a})\label{example:2} Consider the time-invariant version of \eqref{state_process_tv}, \eqref{observation_process_tv} with
\begin{align}
\begin{split}
&A=\diag(1.2, 1.2, 1.2),~C=\begin{bmatrix}
0.8147  &  0.9134 &   0.2785\\
    0.9058  &  0.6324  &  0.5469\\
    0.1270  &  0.0975   & 0.9575
\end{bmatrix},\\
&\Sigma_{\bf w}=\begin{bmatrix}
0.8895  &  1.1744 &   0.2309\\
    1.1744  &  1.8616  &  0.2953\\
    0.2309  &  0.2953  &  0.0614
    \end{bmatrix},~\Sigma_{\bf n}=\diag(1,1,0).
\end{split}\label{inputs}
\end{align}
Clearly, from Lemma \ref{lemma:necessary_suff}, the pair $(A, C)$ is detectable and the pair $(A, \Sigma_{\bf w}^{\frac{1}{2}})$ is stabilizable. Hence the filter ${\bm \xi}_t$ is asymptotically stable,  with
\begin{align}
\bar{\Sigma}=\begin{bmatrix}
2.6928 &  -0.7211 &   0.1847\\
   -0.7211  &  4.0349  &  0.3254\\
    0.1847  &  0.3254  &  0.0645
    \end{bmatrix},\label{noise_matrix}
\end{align}
 and from \eqref{dare} we obtain $\Pi\succ{0}$ which further implies the steady-state solution of $\Sigma\succeq{0}$  both given as follows
\begin{align}
&\Pi=\begin{bmatrix}
6.7910 &  -5.0291  &  0.0798\\
   -5.0291 &   8.9742   & 0.3939\\
    0.0798  &  0.3939  &  0.0714
\end{bmatrix},\label{priori_cov_ss}\\
&\Sigma=\begin{bmatrix}4.0983 &  -4.3080   & -0.1049\\
   -4.3080 &   4.9393  &  0.0684\\
   -0.1049  &  0.0684   & 0.0069
   \end{bmatrix}.\label{posteriori_cov_ss}
\end{align}
We recall using \cite[Corollary 1, Theorem 9]{kostina:2019a}, that the closed form solution of the sum-rate therein under the assumption of {\it uniform rate-distortion allocation} is given by
\begin{align}
R_{[0,\infty],\ind}^{G, KH}(D-\trace(\Sigma))=\frac{p}{2}\log\left(\bar{a}^2+\frac{|\bar{\Sigma}|^{\frac{1}{p}}p}{D-\trace(\Sigma)}\right),\label{sum_rate_kostina_hassibi}
\end{align}
where $\bar{a}\triangleq\abs(|A|)^{\frac{1}{p}}$, $\bar{\Sigma}=\Pi-\Sigma$, with $D>\trace(\Sigma)$. In Fig. \ref{fig:optimal_rates}, we give the optimal numerical solution obtained via Corollary \ref{corollary:complete_optim_numer_infinite_horizon}, (2) using the CVX platform \cite{cvx} and the reverse-waterfilling solution of Theorem \ref{theorem:rev_water} using Algorithm \ref{algo2} (because the input data in \eqref{inputs} satisfy the strong structural properties of Proposition \ref{proposition:strong_structural_properties}, (i)). We compare the optimal sum-rate with the closed-form solution of \eqref{sum_rate_kostina_hassibi}. We observe that the latter is in general highly suboptimal with respect to the optimal numerical solution with the maximum rate-loss (RL), which for this example is approximately ${1.05}$ bits/{vector source}, to be observed at moderate to low rates. A good performance of \eqref{sum_rate_kostina_hassibi} in the sense that it almost coincides with the exact optimal solution can be observed at very high rates. This means that Corollary \ref{corollary:complete_optim_numer_infinite_horizon} and Theorem \ref{theorem:rev_water} that allow non-uniform distortion allocation may achieve significant performance gains compared to \eqref{sum_rate_kostina_hassibi} that only allows uniform distortion allocation. 
\begin{figure}[htp]
\centering
\includegraphics[width=\columnwidth]{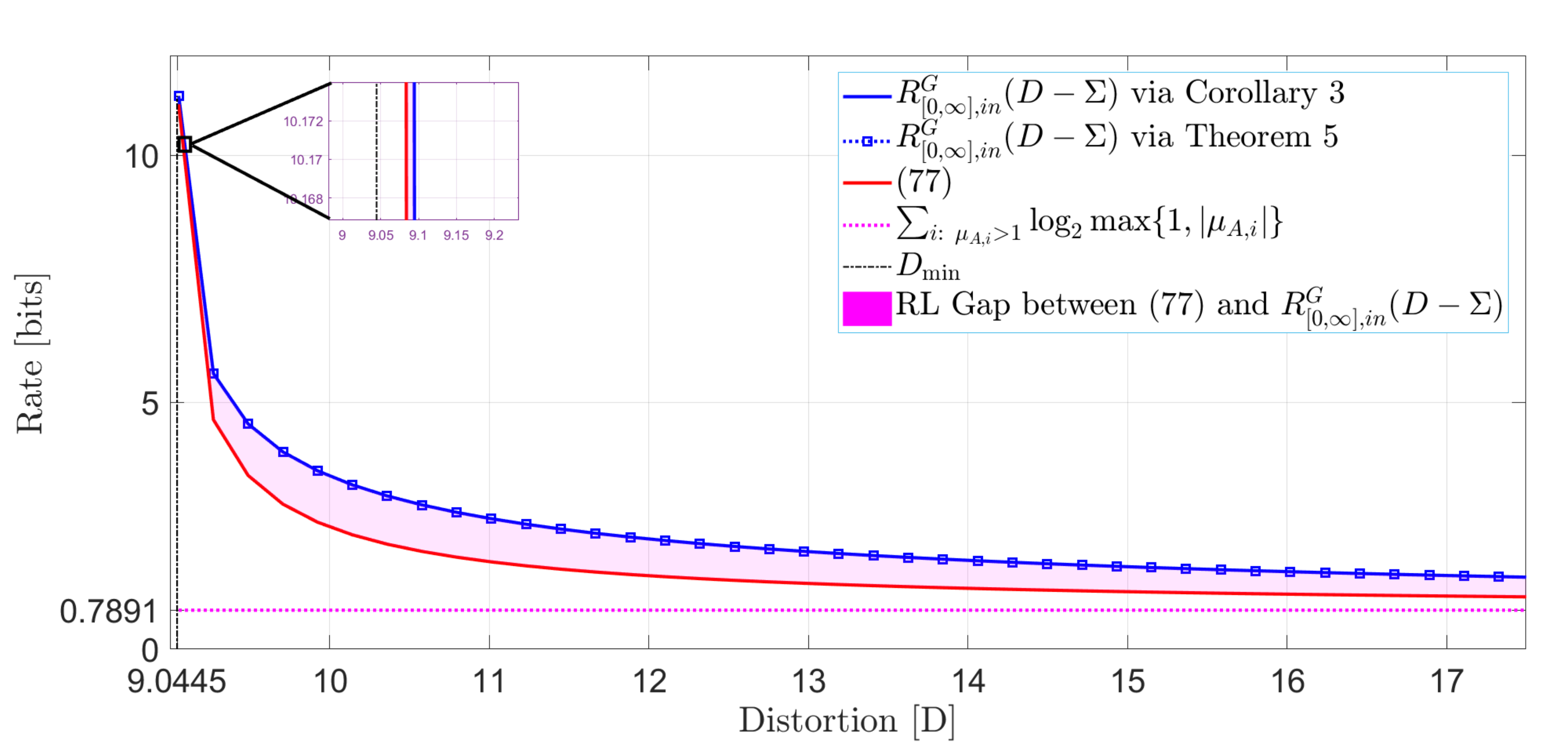}
\caption{{Comparison of the optimal sum-rates obtained via Corollary \ref{corollary:complete_optim_numer_infinite_horizon} and Theorem \ref{theorem:rev_water} with the closed form solution obtained via \eqref{sum_rate_kostina_hassibi}.}}\label{fig:optimal_rates}
\end{figure}
\end{example}
\begin{example}(Convergence to steady-state solution)\label{example:3}
Consider the time-invariant version of \eqref{scalar_noisy_observ_system} with $(\alpha, c, \sigma_{\bf w}^2, \sigma_{\bf n}^2)=(1.1, 0.5, 1, 1),~\forall{t}$. Clearly, from Example \ref{example:1}, the pair $(\alpha, c)$ is detectable and the pair $(\alpha, (\sigma^{2}_{\bf w})^{\frac{1}{2}})$ is stabilizable. Hence the filter ${\bm \xi}_t=\mathbb{E}\{{\bf x}_t|{\bf z}^t\}$ is asymptotically stable and from \eqref{dare_scalar} we obtain $\Pi=3.1215>0$ whereas from \eqref{scalar_dare_posterior} the non-negative solution is $\Sigma=1.7532$. For a given distortion level $D=2.7532>\Sigma$ we obtain via \eqref{closed_form_scalar_inf_hor} $R_{\ind}^G(D-\Sigma)=0.6832$ (bits/source~sample). Using Algorithm \ref{algo1}, we compute \eqref{par_sol_scalar_tv} (normalized over the time horizon $(n+1)$) for sufficiently large time horizon, i.e., $n\longrightarrow{10^{5}}$. In Fig. \ref{fig:steady_state}, we illustrate the asymptotic behavior of Algorithm \ref{algo1} versus (vs) the steady-state solution \eqref{closed_form_scalar_inf_hor} in a semi-logarithmic scale. The two lines are met really fast but do not coincide. In fact, depending on the precision error of Algorithm \ref{algo1} (a reasonable error tolerance is $\epsilon=10^{-9}$) one can also infer about the discrepancy of the two lines. We note that Algorithm \ref{algo1} also gives $D_{[0,n]}^{\min}\approx{\Sigma}$ and $\lim_{t\longrightarrow{10^5}}\sigma^2_{{\bf x}_{t|t-1}}\approx\Pi$. Moreover, the starting point of the plot obtained from Algorithm \ref{algo1} depends on $\sigma^2_{{\bf x}_{0}}$.
\begin{figure}[htp]
\centering
\includegraphics[width=\columnwidth]{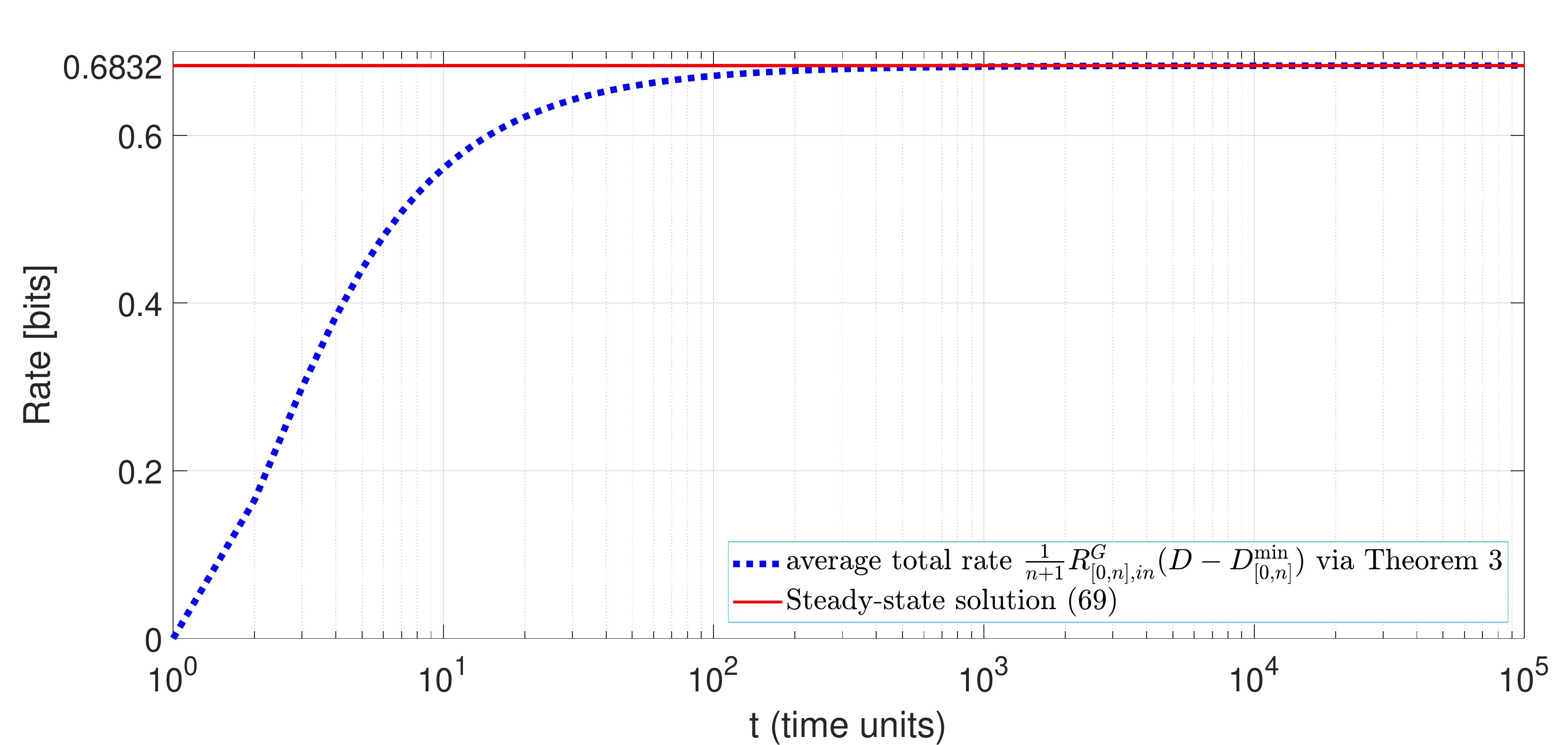}
\caption{{Comparison of Theorem \ref{theorem:opti_num_sol_scalar_tv} vs the steady-state solution of Corollary \ref{corollary:scalar_time_inv} for time-invariant scalar-valued processes.}}\label{fig:steady_state}
\end{figure}
\end{example}
}
{
\section{Conclusions and Ongoing Research}\label{sec:conclusions}
In this paper we revisited the problem of characterizing and computing the indirect NRDF for partially observable multivariate Gauss-Markov processes with hard MSE distortion constraints. We derived the complete characterization and the corresponding optimal test channel realization and gave conditions to ensure existence of solution of the characterization in both finite and infinite time horizon. Moreover, we obtained optimal numerical and closed form solutions for vector and scalar systems under either average total or pointwise MSE distortion constraints. One particularly interested result is the construction of new scalable optimal iterative schemes for time-varying scalar processes and time-invariant multidimensional processes that operate much faster than the standard semidefinite programming algorithms.}
\par {One particular question that we do not address herein but can be further analyzed from our results, is the relaxation of the Gaussian noise process that drives the state of the system model in \eqref{state_process_tv}, \eqref{observation_process_tv} to positive semidefinite covariance matrices. Another important question is the extension of Theorem \ref{theorem:rev_water} to time-varying processes which will require strong time-varying structural properties in the spirit of Proposition \ref{proposition:strong_structural_properties}. Finally, the extension of this problem to controlled processes is also of major importance.}

\appendices

{
\section{Proof of Theorem \ref{theorem:numer_col_finite_time}}\label{proof:theorem:numerical_solution:finite_time}
}
{
{\bf (1)} Under the conditions of the theorem we ensure that there exists an optimal solution for the general characterization of Theorem \ref{theorem:complete_characterization_suff_stat}. Now the objective function in \eqref{complete_characterization_suff_stat} can be reformulated as follows:
\begin{align}
\underbrace{\frac{1}{2}\log|\Sigma_{0|-1}^{\bm \xi}|}_{\mbox{initial time}}+\underbrace{\frac{1}{2}\sum_{t=0}^{n-1}\log\frac{|\Sigma^{\bm \xi}_{t+1|t}|}{|\Sigma^{\bm \xi}_{t|t}|}}_{\mbox{time varying term}}-\underbrace{\frac{1}{2}\log|\Sigma_{n|n}^{\bm \xi}|}_{\mbox{terminal time}}.\label{proof_dec_time_instants}
\end{align}
Note that the first term in \eqref{proof_dec_time_instants} is given whereas the terminal time is decoupled of all previous time instants and can be optimized separately.
Under the assumption that $A_t$ is full rank, the additive time-varying term can be reformulated as follows
\begin{align}
&\frac{1}{2}\sum_{t=0}^{n-1}\log\frac{|\Sigma^{\bm \xi}_{t+1|t}|}{|\Sigma^{\bm \xi}_{t|t}|}=\frac{1}{2}\sum_{t=0}^{n-1}\Big[\log|\Sigma_{t|t}^{\bm \xi}+A_t^{-1}B_{t+1}B_{t+1}\T(A_t\T)^{-1}|\nonumber\\
&\qquad\qquad\qquad+\log\abs(|A_t|)-\log|\Sigma_{t|t}^{\bm \xi}|\Big]\nonumber\\
&=\frac{1}{2}\sum_{t=0}^{n-1}\Big[\log|I_p+(\Sigma_{t|t}^{\bm \xi})^{-\frac{1}{2}}A_t^{-1}B_{t+1}B_{t+1}\T(A_t\T)^{-1}(\Sigma_{t|t}^{\bm \xi})^{-\frac{1}{2}}|\nonumber\\
&\qquad+\log\abs(|A_t|)\Big]\nonumber\\
&\stackrel{(a)}=\frac{1}{2}\sum_{t=0}^{n-1}\Big[\log|I_p+B_{t+1}\T(A_t\T)^{-1}({\Sigma_{t|t}^{\bm \xi}})^{-1}A_t^{-1}B_{t+1}|\nonumber\\
&\qquad+\log\abs(|A_t|)\Big]\nonumber\\
&=\frac{1}{2}\sum_{t=0}^{n-1}\Big[-\log|I_p+B_{t+1}\T(A_t\T)^{-1}({\Sigma_{t|t}^{\bm \xi}})^{-1}A_t^{-1}B_{t+1}|^{-1}\nonumber\\
&\qquad+\log\abs(|A_t|)\Big]
\label{anal_cost_to_go}
\end{align}
where $(a)$ follows from Weinstein-Aronszajn identity \cite[Corollary 18.1.2]{harville:1997}. Due to the monotonicity of the determinant in the first RHS term of \eqref{anal_cost_to_go} and \eqref{proof_dec_time_instants}, we can formulate the optimization problem \eqref{complete_characterization_suff_stat} as follows:
\begin{align}
\begin{split}
&{R}^{G}_{[0,n],\ind}(D-D_{[0,n]}^{\min})= \\
&\min_{\substack{\{\Sigma_{t|t}^{\bm \xi}\succ{0}, \Gamma^{1}_t\succ{0}\}_{t=0}^n\\ 
0\prec\Sigma^{\bm \xi}_{0|0}\preceq\Sigma^{\bm \xi}_{0|-1}\\
0\prec\Sigma^{\bm \xi}_{t+1|t+1}\preceq\Sigma^{\bm \xi}_{t+1|t},~t\in\mathbb{N}_0^{n-1}\\
 \Sigma_{n|n}^{\bm \xi}=\Gamma^{1}_{n}}}\frac{1}{2}\sum_{t=0}^n-\log|\Gamma_t^{1}|+c_1,
\end{split}\label{proof:reform_of_compl_charact}
\end{align}
with the additional LMI constraint $0\prec\Gamma_t^{1}\preceq(I_p+B_{t+1}\T(A_t\T)^{-1}({\Sigma_{t|t}^{\bm \xi}})^{-1}A_t^{-1}B_{t+1})^{-1},~t\in\mathbb{N}_0^{n-1}$, where $c_1=\frac{1}{2}\log|\Sigma_{0|-1}^{\bm \xi}|+\sum_{t=0}^{n-1}\log\abs\left(|A_t|\right)$. Note that the equality constraint $\Sigma_{n|n}^{\bm \xi}=\Gamma_n^1$ in \eqref{proof:reform_of_compl_charact} follows because when $t=n$ at the objective function we only optimize with respect to $-\log|\Sigma_{n|n}|$ which has been decoupled from the previous time instants therefore the use of the additional ``slack'' variable $\Gamma_n^{1}$ is not needed and we simply take the equality constraint. Now using Woodbury matrix identity \cite[Theorem 18.2.8]{harville:1997} in the additional LMI constraint we obtain
\begin{align}
0\prec\Gamma_t^{1}\preceq{I}_p-B_{t+1}\T(\Sigma_{t+1|t}^{\bm \xi})^{-1}B_{t+1},\label{reform_woodbury_matrix_identity}
\end{align}
where $\Sigma_{t+1|t}^{\bm \xi}=A{\Sigma_{t|t}^{\bm \xi}}A\T+B_{t+1}B_{t+1}\T$, which is equivalent (as a Schur complement) to the constraint block matrix in \eqref{numer_sol_suff_stat:finite_time_1}. \\
{\bf (2)} This follows similar to {\bf (1)} hence we omit it. This completes the derivation.} 

{
\section{Proof of Theorem \ref{theorem:opti_num_sol_scalar_tv}}\label{proof:theorem:opti_num_sol_scalar_tv}
To solve the problem, we employ Karush-Kuhn-Tucker (KKT) conditions \cite[Chapter 5.5.3]{boyd:2004} which are for the convex program in \eqref{complete_characterization_suff_stat_scalar} necessary and sufficient conditions for global optimality. Similar to the proof of Theorem \ref{theorem:numer_col_finite_time}, we reformulate the objective in \eqref{complete_characterization_suff_stat_scalar} to decouple the dependency of $\{\sigma_{{\bm \xi}_{t|t}}^2:~t\in\mathbb{N}_0^n\}$, from previous $\{\sigma_{{\bm \xi}_{t|t}}^2:~t\in\mathbb{N}_0^n\}$ at each instant of time:
\begin{align}
&\frac{1}{2}\sum_{t=0}^n\log\left(\frac{\sigma_{{\bm \xi}_{t|t-1}}^2}{\sigma_{{\bm \xi}_{t|t}}^2}\right)=\nonumber\\
&\frac{1}{2}\left[\underbrace{\log(\sigma_{{\bm \xi}_{0|-1}}^2)}_{given}+\sum_{t=0}^{n-1}\log\left(\alpha_t^2+\frac{\sigma_{{\upsilon}_{t+1}}^2}{\sigma_{{\bm \xi}_{t|t}}^2}\right)-\log(\sigma_{{\bm \xi}_{n|n}}^2)\right].
\label{proof:dec_obj_scalar_tv}
\end{align}
Introduce the augmented Lagrangian as follows:
\begin{align}
&J(\{\sigma^2_{{\bm \xi}_{t|t}}, \lambda_t,~\phi_t\}_{t=0}^n, \theta)=\frac{1}{2}\Biggl[\log(\sigma_{{\bm \xi}_{0|-1}}^2)-\log(\sigma_{{\bm \xi}_{n|n}}^2)\nonumber\\
&+\sum_{t=0}^{n-1}\log\left(\alpha_t^2+\frac{\sigma_{{\upsilon}_{t+1}}^2}{\sigma_{{\bm \xi}_{t|t}}^2}\right)\Biggr]+\sum_{t=0}^n\phi_t\left(\sigma_{{\bm \xi}_{t|t}}^2-\sigma_{{\bm \xi}_{t|t-1}}^2\right)\nonumber\\
&+\theta\left(\sum_{t=0}^n\sigma_{{\bm \xi}_{t|t}}^2-(n+1)(D-D_{[0,n]}^{\min})\right)-\sum_{t=0}^n\lambda_t\sigma_{{\bm \xi}_{t|t}}^2,\label{augmented_lagrangian_scalar_tv}
\end{align}
where $\theta\geq{0}$, $\lambda_t\geq{0},~\phi_t\geq{0},~\forall{t}$. The KKT conditions are as follows:
\begin{align}
&\frac{J(\{\sigma^2_{{\bm \xi}_{t|t}}, \lambda_t,~\phi_t\}_{t=0}^n, \theta)}{\partial\sigma_{{\bm \xi}_{t|t}}^2}\Biggr{|_{\substack{\sigma_{{\bm \xi}_{t|t}}^2=\sigma_{{\bm \xi}_{t|t}}^{2,*}\\
\theta=\theta^*\\
\lambda_t=\lambda_t^{*}\\
\phi_t=\phi_t^{*}}}}=0,\label{stationarity_cond_tv}\\
&\theta\left(\sum_{t=0}^n\sigma_{{\bm \xi}_{t|t}}^2-(n+1)(D-D_{[0,n]}^{\min})\right)=0,~\mu_t\sigma_{{\bm \xi}_{t|t}}^2=0,\label{complementary_slackness_1_tv}\\
&\phi_t\left(\sigma_{{\bm \xi}_{t|t}}^2-\sigma_{{\bm \xi}_{t|t-1}}^2\right)=0,
\label{complementary_slackness_2_tv}\\
&\sum_{t=0}^n\sigma_{{\bm \xi}_{t|t}}^2\geq(n+1)(D-D_{[0,n]}^{\min}),~\sigma_{{\bm \xi}_{t|t}}^2\geq{0},~\sigma_{{\bm \xi}_{t|t-1}}^2\geq\sigma_{{\bm \xi}_{t|t}}^2,\label{primal_feasibility_cond_tv}\\
&\theta\geq{0},~\lambda_t\geq{0},~\phi_t\geq{0},\forall{t},\label{dual_feasibility_cond_tv}
\end{align} 
where \eqref{stationarity_cond_tv} is the first order partial derivative test; \eqref{complementary_slackness_1_tv}, \eqref{complementary_slackness_2_tv} are the complementary slackness conditions; \eqref{primal_feasibility_cond_tv} are the primal feasibility conditions and \eqref{dual_feasibility_cond_tv} are the dual feasibility conditions.\\
Next, we check the conditions. First observe that, by definition, $\lambda_t=\lambda_t^*=0$ because $\sigma^2_{{\bm \xi}_{t|t}}>0,~\forall{t}$. Moreover $\phi_t=\phi_t^*=0,~\forall{t}$, because  only then we have positive rates. In other words, if for some $t$ $\phi_t^*>0$, then the rate is zero and it can be excluded from the optimal solution of the total rates. Furthermore, $\theta=\theta^*>0$ because by the convexity of the problem and the complementary slackness conditions \eqref{complementary_slackness_1_tv} the distortion constraint holds with equality (the solution occurs on the boundary). Next, we solve \eqref{stationarity_cond_tv} at each instant of time as follows:
\begin{align}
&\frac{J(\{\sigma^2_{{\bm \xi}_{t|t}}, 0,~0\}_{t=0}^n, \theta)}{\partial\sigma_{{\bm \xi}_{t|t}}^2}\Biggr{|_{\substack{\sigma_{{\bm \xi}_{t|t}}^2=\sigma_{{\bm \xi}_{t|t}}^{2,*}\\
\theta=\theta^*}}}\nonumber\\
&=\frac{1}{2}\left(-\frac{\sigma^2_{{\upsilon}_{t+1}}}{\sigma^2_{{\bm \xi}_{t|t}}\left(\alpha_t^2\sigma^2_{{\bm \xi}_{t|t}}+\sigma^2_{{\upsilon}_{t+1}}\right)}\right)+\theta=0,~t\in\mathbb{N}_0^{n-1}\label{stationarity_tv_1}\\
&\frac{J(\{\sigma^2_{{\bm \xi}_{t|t}}, 0,~0\}_{t=0}^n, \theta)}{\partial\sigma_{{\bm \xi}_{n|n}}^2}\Biggr{|_{\substack{\sigma_{{\bm \xi}_{t|t}}^2=\sigma_{{\bm \xi}_{t|t}}^{2,*}\\
\theta=\theta^*}}}\nonumber\\
&=\frac{1}{2}\left(-\frac{1}{\sigma_{{\bm \xi}_{n|n}}^{2,*}}\right)+\theta=0\Longrightarrow\sigma_{{\bm \xi}_{n|n}}^{2,*}=\frac{1}{2\theta},~t=n.\label{stationarity_tv_2}
\end{align}
 The solution of \eqref{stationarity_tv_1} results into a quadratic equation with one positive and one negative solution. By definition, we choose the positive solution given in \eqref{sol_kkt_tv} whereas at $t=n$ the solution is given by \eqref{stationarity_tv_2}. The non-negativity of the objective function in \eqref{par_sol_scalar_tv} is guaranteed via \eqref{rev_water_scalar_tv}. This completes the proof.
}
{
\section{Proof of Theorem \ref{theorem:existence_asymptotic_limit}}\label{proof:theorem:existence_asymptotic_limit}
First note that under the conditions of the theorem, we have the unique stabilizing solution $\lim_{t\longrightarrow\infty}\Pi_t=\Pi\succ{0}$ and consequently $\lim_{t\longrightarrow\infty}\Sigma_t=\Sigma\succeq{0}$. This in turn implies via \eqref{KF_classical} of Lemma \ref{lemma:classical_kf} that  $\lim_{t\longrightarrow\infty}\bar{\Sigma}_t=\bar{\Sigma}$. The specific steady state solution corresponds to an asymptotically stable filter. Then, the objective function in \eqref{complete_characterization_suff_stat_infinite_horizon} is obtained as follows
\begin{align}
&\limsup_{n\longrightarrow\infty}\frac{1}{n+1}\sum_{t=0}^n\log\frac{|\Sigma^{\bm \xi}_{t|t-1}|}{|\Sigma_{t|t}^{\bm \xi}|}\nonumber\\
&\stackrel{(a)}=\limsup_{n\longrightarrow\infty}\frac{1}{n+1}\sum_{t=0}^n\log\frac{|A\Sigma^{\bm \xi}_{t-1|t-1}A\T+\bar{\Sigma}_t|}{|\Sigma_{t|t}^{\bm \xi}|}\label{asympt_limit_proof}\\
&\stackrel{(b)}=\eqref{complete_characterization_suff_stat_infinite_horizon},\nonumber 
\end{align}
where $(a)$ follows because $\Sigma_{t|t-1}^{\bm \xi}={A}\Sigma^{\bm \xi}_{t-1|t-1}{A}\T+\bar{\Sigma}_t$;  $(b)$  follows because we restrict the numerator and denominator in \eqref{asympt_limit_proof} to be have a time invariant value  (because we impose the optimal minimizer to be time invariant and the corresponding output distribution to be time-invariant with a unique invariant distribution). Note that $\Pi^{\xi}$ is given by \eqref{a_priori_tinv} and $\{\bar{\Sigma}_n:~n\in\mathbb{N}_0\}$ is a convergent sequence (by the conditions of the theorem) and its steady-state (time invariant) solution is $\bar{\Sigma}=\lim_{t\longrightarrow\infty}\Sigma_n$. The constraint set in \eqref{complete_characterization_suff_stat_infinite_horizon} is obtained because via Remark \ref{remark:suff_cond_characterization} we ensure a finite solution to the optimization problem if we impose the strict LMI $0\prec\Sigma^{\bm \xi}\preceq\Pi^{\bm \xi}$ which implies that $\Sigma^{\bm \xi}\succ{0}$ and $\Pi^{\bm \xi}\succ{0}$. From the conditions of the theorem, we have a convergent sequence $\{\Sigma_n:~n\in\mathbb{N}_0\}$, i.e.,~$\lim_{n\longrightarrow\infty}\Sigma_n=\Sigma$ which further means that $\{\trace(\Sigma_n):~n\in\mathbb{N}_0\}$ is also convergent. This in turn implies that $\frac{1}{n+1}\sum_{t=0}^n\trace(\Sigma_t)=\trace(\Sigma)$ as $n\longrightarrow\infty$ which is precisely \eqref{steady_state_min_dist}. This completes the characterization of \eqref{complete_characterization_suff_stat_infinite_horizon}. The optimal time-invariant test channel realization \eqref{optimal_realization_suff_stat_infinite_horizon} follows easily from the conditions of the theorem. This completes the derivation.
}

{
\section*{Acknowledgements}
The authors wish to thank the Associate Editor and the anonymous reviewers for their valuable
comments and suggestions. We are especially indebted to one the reviewers who recognized that the problem studied in the revised form of this paper was still an open problem in the literature. 
}

\bibliographystyle{IEEEtran}
\bibliography{string,literature_conf}

\begin{thebibliography}{10}
\providecommand{\url}[1]{#1}
\csname url@samestyle\endcsname
\providecommand{\newblock}{\relax}
\providecommand{\bibinfo}[2]{#2}
\providecommand{\BIBentrySTDinterwordspacing}{\spaceskip=0pt\relax}
\providecommand{\BIBentryALTinterwordstretchfactor}{4}
\providecommand{\BIBentryALTinterwordspacing}{\spaceskip=\fontdimen2\font plus
\BIBentryALTinterwordstretchfactor\fontdimen3\font minus
  \fontdimen4\font\relax}
\providecommand{\BIBforeignlanguage}[2]{{%
\expandafter\ifx\csname l@#1\endcsname\relax
\typeout{** WARNING: IEEEtran.bst: No hyphenation pattern has been}%
\typeout{** loaded for the language `#1'. Using the pattern for}%
\typeout{** the default language instead.}%
\else
\language=\csname l@#1\endcsname
\fi
#2}}
\providecommand{\BIBdecl}{\relax}
\BIBdecl

\bibitem{gorbunov:1972}
A.~K. Gorbunov and M.~S. Pinsker, ``Nonanticipatory and prognostic epsilon
  entropies and message generation rates,'' \emph{Problems Inf. Transmiss.},
  vol.~9, no.~3, pp. 184--191, {J}uly-{S}ept. 1972.

\bibitem{gorbunov:1972b}
------, ``Prognostic epsilon entropy of a {G}aussian message and a {G}aussian
  source,'' \emph{Problems Inf. Transmiss.}, vol.~10, no.~2, pp. 93--109,
  {A}pr.-{J}une 1972, translation from Problemy Peredachi Informatsii, vol. 10,
  no. 2, pp. 5-–25, April-June 1974.

\bibitem{derpich:2012}
M.~S. Derpich and J.~{\O}stergaard, ``Improved upper bounds to the causal
  quadratic rate-distortion function for {G}aussian stationary sources,''
  \emph{IEEE Trans. Inf. Theory}, vol.~58, no.~5, pp. 3131 -- 3152, May 2012.

\bibitem{stavrou:2018}
P.~A. {Stavrou}, J.~{\O}stergaard, and C.~D. {Charalambous}, ``Zero-delay rate
  distortion via filtering for vector-valued {G}aussian sources,'' \emph{IEEE
  J. Sel. Topics Signal Process.}, vol.~12, no.~5, pp. 841--856, Oct 2018.

\bibitem{tatikonda:2004}
S.~Tatikonda, A.~Sahai, and S.~Mitter, ``Stochastic linear control over a
  communication channel,'' \emph{IEEE Trans. Autom. Control}, vol.~49, pp. 1549
  -- 1561, 2004.

\bibitem{charalambous:2014}
C.~D. {Charalambous}, P.~A. {Stavrou}, and N.~U. {Ahmed}, ``Nonanticipative
  rate distortion function and relations to filtering theory,'' \emph{IEEE
  Transactions on Automatic Control}, vol.~59, no.~4, pp. 937--952, 2014.

\bibitem{cover-thomas:2006}
T.~M. Cover and J.~A. Thomas, \emph{Elements of Information Theory},
  2nd~ed.\hskip 1em plus 0.5em minus 0.4em\relax John Wiley \& Sons, Inc.,
  Hoboken, New Jersey, 2006.

\bibitem{kostina:2019a}
V.~{Kostina} and B.~{Hassibi}, ``Rate-cost tradeoffs in control,'' \emph{IEEE
  Trans. Autom. Control}, vol.~64, no.~11, pp. 4525--4540, Nov 2019.

\bibitem{stavrou:2020tac}
P.~A. {Stavrou}, T.~{Tanaka}, and S.~{Tatikonda}, ``The time-invariant
  multidimensional {G}aussian sequential rate-distortion problem revisited,''
  \emph{IEEE Transactions on Automatic Control}, vol.~65, no.~5, pp.
  2245--2249, 2020.

\bibitem{tanaka:2017}
T.~Tanaka, K.~K.~K. Kim, P.~A. Parrilo, and S.~K. Mitter, ``Semidefinite
  programming approach to {G}aussian sequential rate-distortion trade-offs,''
  \emph{IEEE Trans. Autom. Control}, vol.~62, no.~4, pp. 1896--1910, April
  2017.

\bibitem{stavrou:2018siam}
P.~A. Stavrou, T.~Charalambous, C.~D. Charalambous, and S.~Loyka, ``Optimal
  estimation via nonanticipative rate distortion function and applications to
  time-varying {G}auss-{M}arkov processes,'' \emph{SIAM J. on Control Optim.},
  vol.~56, no.~5, pp. 3731--3765, 2018.

\bibitem{fuglsig:2019}
A.~J. Fuglsig and J.~{\O}stergaard, ``Zero-delay multiple descriptions of
  stationary scalar {G}auss-{M}arkov sources,'' \emph{Entropy}, vol.~21,
  no.~12, 2019.

\bibitem{charalambous:2019cdc}
C.~D. {Charalambous}, C.~{Kourtellaris}, T.~{Charalambous}, and J.~H. {van
  Schuppen}, ``Generalizations of nonanticipative rate distortion function to
  multivariate nonstationary {G}aussian autoregressive processes,'' in
  \emph{2019 IEEE 58th Conference on Decision and Control (CDC)}, 2019, pp.
  8190--8195.

\bibitem{tanaka:2018}
T.~{Tanaka}, P.~M. {Esfahani}, and S.~K. {Mitter}, ``{LQG} control with minimum
  directed information: Semidefinite programming approach,'' \emph{IEEE Trans.
  Autom. Control}, vol.~63, no.~1, pp. 37--52, Jan 2018.

\bibitem{tanaka:2015}
T.~{Tanaka}, ``Zero-delay rate-distortion optimization for partially observable
  {G}auss-{M}arkov processes,'' in \emph{Proc. IEEE Conf. Decision Control},
  Dec 2015, pp. 5725--5730.

\bibitem{massey:1990}
J.~L. Massey, ``Causality, feedback and directed information,'' in \emph{Proc.
  Int. Symp. Inf. Theory Appl.}, {N}ov. 27-30 1990, pp. 303--305.

\bibitem{stavrou:2021tac}
P.~A. Stavrou and M.~Skoglund, ``Asymptotic reverse-waterfilling algorithm for
  certain classes of vector {G}auss-{M}arkov processes,'' \emph{IEEE Trans.
  Autom. Control}, pp. 1--1, 2021.

\bibitem{stavrou:2020entropy}
P.~A. Stavrou, J.~{\O}stergaard, and M.~Skoglund, ``Bounds on the sum-rate of
  {MIMO} causal source coding systems with memory under spatio-temporal
  distortion constraints,'' \emph{Entropy}, vol.~22, no.~8, 2020.

\bibitem{dupuis-ellis:1997}
P.~Dupuis and R.~S. Ellis, \emph{A Weak Convergence Approach to the Theory of
  Large Deviations}.\hskip 1em plus 0.5em minus 0.4em\relax John Wiley \& Sons,
  Inc., New York, 1997.

\bibitem{dobrushin:1962}
R.~{Dobrushin} and B.~{Tsybakov}, ``Information transmission with additional
  noise,'' \emph{IRE Trans. Info. Theory}, vol.~8, no.~5, pp. 293--304, Sep.
  1962.

\bibitem{wolf-ziv:1970}
J.~Wolf and J.~Ziv, ``Transmission of noisy information to a noisy receiver
  with minimum distortion,'' \emph{IEEE Trans. Inf. Theory}, vol.~16, no.~4,
  pp. 406--411, 1970.

\bibitem{berger:1971}
T.~Berger, \emph{Rate Distortion Theory:~A Mathematical Basis for Data
  Compression}.\hskip 1em plus 0.5em minus 0.4em\relax Englewood Cliffs, NJ:
  Prentice-Hall, 1971.

\bibitem{witsenhausen:1980}
H.~{Witsenhausen}, ``Indirect rate distortion problems,'' \emph{IEEE Trans.
  Inf. Theory}, vol.~26, no.~5, pp. 518--521, Sep. 1980.

\bibitem{charalambous:2016}
C.~D. Charalambous and P.~A. Stavrou, ``Directed information on abstract
  spaces: Properties and variational equalities,'' \emph{IEEE Trans. Inf.
  Theory}, vol.~62, no.~11, pp. 6019--6052, Nov 2016.

\bibitem{ihara1993}
S.~Ihara, \emph{Information theory - for Continuous Systems}.\hskip 1em plus
  0.5em minus 0.4em\relax World Scientific, 1993.

\bibitem{anderson:1979}
B.~D.~O. Anderson and J.~B. Moore, \emph{{Optimal Filtering}}.\hskip 1em plus
  0.5em minus 0.4em\relax Englewood Cliffs, NJ: Prentice-Hall, 1979.

\bibitem{kailath:2000}
T.~Kailath, A.~H. Sayed, and B.~Hassibi, \emph{Linear Estimation}.\hskip 1em
  plus 0.5em minus 0.4em\relax Upper Saddle River, New Jersey: Prentice Hall,
  200.

\bibitem{simon:2006}
D.~Simon, \emph{Optimal State Estimation: {K}alman, $H_{\infty}$, and Nonlinear
  Approaches}.\hskip 1em plus 0.5em minus 0.4em\relax Wiley-Interscience, 2006.

\bibitem{vanschuppen:2021}
J.~V. Schuppen, \emph{Control and Systems Theory of Discrete-Time Stochastic
  Systems}, ser. Communications and Control Engineering.\hskip 1em plus 0.5em
  minus 0.4em\relax Springer, 2021.

\bibitem{charalambous2020new}
C.~D. Charalambous, C.~Kourtellaris, and S.~Louka, ``New formulas of feedback
  capacity for {AGN} channels with memory: A time-domain sufficient statistic
  approach,'' 2020.

\bibitem{bertsekas:2005}
D.~P. Bertsekas, \emph{Dynamic programming and optimal control}.\hskip 1em plus
  0.5em minus 0.4em\relax Athena Scientific, 2005.

\bibitem{barnett:1990}
S.~Barnett, \emph{Matrices: Methods and Applications}.\hskip 1em plus 0.5em
  minus 0.4em\relax Oxford University Press, 1990.

\bibitem{stavrou:2021tac2}
P.~A. Stavrou, M.~Skoglund, and T.~Tanaka, ``Sequential source coding for
  stochastic systems subject to finite rate constraints,'' \emph{IEEE
  Transactions on Automatic Control}, pp. 1--1, 2021.

\bibitem{stavrou:2018lcss}
P.~A. Stavrou, T.~Charalambous, and C.~D. Charalambous, ``Finite-time
  nonanticipative rate distortion function for time-varying scalar-valued
  {G}auss-{M}arkov sources,'' \emph{IEEE Control Syst. Lett.}, vol.~2, no.~1,
  pp. 175--180, Jan 2018.

\bibitem{harville:1997}
D.~A. Harville, \emph{Matrix Algebra From a Statistician's Perspective}.\hskip
  1em plus 0.5em minus 0.4em\relax Springer-Verlag, 1997.

\bibitem{zhang:2011}
F.~Zhang, \emph{Matrix Theory: Basic results and techniques}, 2nd~ed.\hskip 1em
  plus 0.5em minus 0.4em\relax Springer-Verlag New York, 2011.

\bibitem{alman-williams:2021}
J.~Alman and V.~V. Williams, \emph{A Refined Laser Method and Faster Matrix
  Multiplication}, ser. Proceedings of the 2021 ACM-SIAM Symposium on Discrete
  Algorithms (SODA), 2021, pp. 522--539.

\bibitem{kostina:2020}
\BIBentryALTinterwordspacing
V.~{Kostina}, ``Fundamental limitations in distributed tracking,'' in
  \emph{IEEE International Symposium on Information Theory}, 2020. [Online].
  Available: \url{https://arxiv.org/abs/1910.02534v1}
\BIBentrySTDinterwordspacing

\bibitem{cvx}
M.~Grant and S.~Boyd, ``{CVX}: Matlab software for disciplined convex
  programming, version 2.1,'' \url{http://cvxr.com/cvx}, Mar. 2014.

\bibitem{boyd:2004}
S.~Boyd and L.~Vandenberghe, \emph{Convex Optimization}.\hskip 1em plus 0.5em
  minus 0.4em\relax New York, NY, USA: Cambridge University Press, 2004.

\end{thebibliography}

%
%
%

\begin{IEEEbiography}
{Photios A. Stavrou} (S'10-M'16) received his D. Eng in 2008 from the Department of Electrical and Computer Engineering (ECE) of the Faculty of Engineering at Aristotle University of Thessaloniki, Greece and his Ph.D degree in 2016 from the Department of ECE of the Faculty of Engineering at University of Cyprus, Cyprus. From November of 2016 to October of 2017, he was a post-doctoral researcher at the Department of Electronic Systems at Aalborg University, Denmark. From November of 2017 to October 2019 he is a post-doctoral researcher at the Division of Information Science and Engineering at KTH Royal Institute of Technology, Sweden. As of November of 2019 he is a researcher at the same institution. His research interests span information and communication theories, communication for networked control systems, feedback and privacy in communication, optimization and state estimation.
\end{IEEEbiography}

\begin{IEEEbiography}
{Mikael Skoglund}
(S'93-M'97-SM'04-F'19) received the Ph.D.~degree in 1997 from Chalmers University of Technology, Sweden.  In 1997, he joined
the Royal Institute of Technology (KTH), Stockholm, Sweden, where he was appointed to the Chair in Communication Theory in 2003.  At KTH he heads the Division of Information Science and Engineering, and the Department of Intelligent Systems. Dr.~Skoglund has worked on problems in source-channel coding, coding and transmission for wireless communications, Shannon theory, information and control, and statistical signal processing. He has authored and co-authored some 160 journals and 380 conference papers. Dr.~Skoglund is a Fellow of the IEEE. During 2003--08 he was an associate editor for the IEEE Transactions on Communications and during 2008--12 he was on the editorial board for the IEEE Transactions on Information Theory. He has served on numerous technical program committees for IEEE sponsored conferences, he was general co-chair for IEEE ITW 2019, and he will serve as TPC co-chair for IEEE ISIT 2022.
\end{IEEEbiography}

\end{document}